\documentclass[12pt]{JHEP3}
\def\a{\alpha}\def\b{\beta}\def\g{\gamma}\def\d{\delta}\def\e{\epsilon }
\def\k{\kappa}

\title{
 Superfield Theories in
Tensorial Superspaces and the Dynamics of Higher Spin Fields}

\bigskip
\author
{I. Bandos$^{\dagger \ddagger}$, P. Pasti$^{\ast}$, D.
Sorokin$^{\ast}$ and Mario Tonin$^{\ast}$
\\
\\
  {\small $^{\dagger}$ \it Departamento de F\'{\i}sica Te\'orica and IFIC,
 46100-Burjassot (Valencia), Spain}
\\
{\small $^{\ast}$ \it Dipartimento di Fisica ``Galileo Galilei",
Universit\'{a} degli Studi di Padova}
\\
{\small ${\&}$ \it INFN, Sezioni di Padova, 35131, Padova, Italia}
\\
{\small $^{\ddagger}$ \it Institute for Theoretical Physics, NSC
KIPT, 61108, Kharkov, Ukraine}
\\{} \\ {\rm Preprint} {\bf FTUV-04-2107}, {\rm arXiv:}{\bf hep-th/0407180}}

\abstract{ We present the superfield generalization of free higher
spin equations in tensorial superspaces and analyze tensorial
supergravities with $GL(n)$ and $SL(n)$ holonomy  as a possible
framework for the construction of a non--linear higher spin field
theory. Surprisingly enough, we find that the most general
solution of the supergravity constraints is given by a class of
superconformally flat and $OSp(1|n)$--related geometries. Because
of the conformal symmetry of the supergravity constraints and of
the higher spin field equations such geometries turn out to be
trivial in the sense that they cannot generate a `minimal'
coupling of higher spin fields to their potentials even in curved
backgrounds with a non--zero cosmological constant. This suggests
that the construction of interacting higher spin theories in this
framework might require an extension of the tensorial superspace
with additional coordinates such as twistor--like spinor variables
which are used to construct the $OSp(1|2n)$ invariant (`preonic')
superparticle action in tensorial superspace.}

\begin{document}

\maketitle

\def\theequation{\arabic{section}.\arabic{equation}}
\setcounter{equation}0
\section{Introduction}

The problem of self--consistent interactions of higher spin fields
is one of the longstanding problems of theoretical physics (see
\cite{V01} for references, \cite{d} for an elementary review and
\cite{Misha0304b} for recent progress). It is known that higher
spin fields can consistently interact in a space--time with a
non--vanishing cosmological constant. The gravitational
interactions of the fermionic fields require the space--time to be
of an anti--de--Sitter type \cite{fradkin+vasiliev,Vasiliev89}.
The interactions should simultaneously involve an infinite set of
fields of an arbitrary high spin and their higher derivatives
\cite{bengtssonal,Berendsal,fradkin+vasiliev,deserD,Vasiliev89}.

Several powerful methods have been developed to deal with theories
which contain an infinite tower of higher spin fields. In
particular, the star product formalism was used to construct
higher spin theories \cite{V94,V99} even earlier than it was
applied to the study of effects of non--commutativity in String
Theory \cite{SW99}. Actually, String Theory itself contains an
infinite tower of interacting massive higher spin excitations. In
a tensionless string limit the higher spin modes become massless
and in a linear approximation satisfy free higher spin equations
of motion (see e.g. \cite{limit} and references therein for more
details).\footnote{Note that these papers deal with a tensionless
limit of  ordinary (super)strings which differs from tensionless
or so called null (super)strings \cite{BZnull}.} However a much
more non--trivial problem is to extract from the string effective
action the information about the structure of higher spin
interactions.

In \cite{fronsdal1} Fronsdal proposed another way of formulating
higher--spin field theory. He conjectured that four--dimensional
higher spin field theory can be realized as a field theory on a
ten--dimensional `tensorial' manifold parametrized by the
coordinates
\begin{equation}\label{x}
X^{\alpha\beta}=X^{\beta\alpha}={1\over
2}x^m\gamma_m^{\alpha\beta}+{1\over 4}
y^{mn}\gamma_{mn}^{\alpha\beta},
\end{equation}
$$
\quad m,n=0,1,2,3\,; \quad \alpha,\beta=1,2,3,4\,,
$$
 where $x^m$ are associated with four
coordinates of the conventional $D=4$ space--time and six
$y^{mn}=-y^{mn}$ describe spin degrees of freedom.

The assumption was that by analogy with, for example $D=10$ or
$D=11$ supergravities,  there may exist a theory in a
ten--dimensional space whose alternative Kaluza--Klein reduction
may lead in $D=4$ to an infinite tower of fields with increasing
spins instead of the infinite tower of Kaluza--Klein particles of
increasing mass. It was argued that the symmetry group of the
theory should be $OSp(1|8)\supset SU(2,2)$, which contains the
$D=4$ conformal group as a subgroup such that an irreducible
(oscillator) representation of $OSp(1|8)$ contains each and every
massless higher spin representation of $SU(2,2)$ only once. So the
idea was that using a single representation of $OSp(1|8)$ in the
ten--dimensional tensorial space one could describe an infinite
tower of higher spin fields in $D=4$ space--time.

This proposal (rather accidentally) found its dynamical
realization in the $OSp(1|2n)$--invariant model of a twistor
superparticle propagating in a flat tensorial superspace
($X^{\alpha\beta}=X^{\beta\alpha}, \theta^\alpha$)
($\alpha,\beta=1,\cdots, n$, with $n=4$ corresponding to the
Fronsdal case (\ref{x})) \cite{BL98,BL98'}. The quantization of
this model was shown \cite{BLS99} to produce the infinite tower of
free massless fields of all possible spins in $D=4$ space--time
and an infinite set of higher spin fields in higher dimensions. In
the general case the bosonic dimension of the tensorial superspace
is ${{n(n+1)}\over 2}$. In particular, the case $n=32$ has been
considered (see \cite{csg}) as a point--like model for a BPS
preon, a hypothetical constituent of M--theory \cite{BPSpreon}.

The superparticle action in the flat tensorial superspace has the
following form
\begin{eqnarray}\label{S0}
S = \int d \tau \, [\dot{X}^{\alpha\beta}(\tau) - i
\dot{\theta}^{\alpha}(\tau)
{\theta}^{\beta}(\tau)]\lambda_{\alpha}\lambda_{\beta}\; ,
\end{eqnarray}
where $\lambda_\alpha(\tau)$ are auxiliary commuting spinor
variables. From (\ref{S0}) it follows that the particle momentum
is $P_{\alpha\beta}=\lambda_\alpha\lambda_\beta$, which in the
tensorial spaces associated with $4,6$ and 10--dimensional
space--times implies that the quantum states of the superparticle
are massless \cite{BL98,BL98'}. Note that this is the direct
analog and generalization of the Cartan--Penrose (twistor)
realization of the light--like momentum of massless states.

The action (\ref{S0}) is non--manifestly invariant under the rigid
transformations of $OSp(1|2n)$
\cite{BL98,BL98',BLS99,V01s,V01c,Dima} but is manifestly invariant
under the transformations of its subgroup $GL(n)$ acting on $X$,
$\theta$ and $\lambda$ as follows
\begin{equation}\label{gl}
X'^{\alpha\beta}=X^{\alpha'\beta'}G_{\alpha'}^{~~\alpha}G_{\beta'}^{~~\beta}\,,
\quad \theta'^{\alpha}=\theta^{\alpha'}G_{\alpha'}^{~~\alpha},
\quad \lambda'_\alpha=G_\alpha^{-1\alpha'}\lambda_{\alpha'}\,.
\end{equation}

Superparticle models and free field theories in flat tensorial
superspaces and on supergroup manifolds $OSp(1|n)$ have been
studied in detail   in
\cite{BLS99,BLPS,V01s,V01c,Misha,Dima,Dima03,V03}. It was
conjectured in \cite{BLPS,V01s} and shown in \cite{Misha,Dima}
that a field theory on $OSp(1|4)$ is classically equivalent to the
$OSp(1|8)$--invariant free higher spin field theory in $AdS_4$.

Interestingly enough, the spectrum of the quantum states and the
wave equations which one obtains by quantizing the particle
propagating in the {\sl bosonic} tensorial space is {\sl
supersymmetric} and possesses $OSp(1|2n)$ symmetry \cite{V01s},
while the spectrum of the quantum states of the particle
propagating in tensorial {\sl super}space is the doubly degenerate
spectrum of the `bosonic' tensorial particle \cite{BLS99}.

In the `bosonic' case (i.e. when $\theta^\alpha=0$) the
quantization of the model (\ref{S0}) results in the following
equation of motion of the particle wave function
$\Phi(X^{\alpha\beta},\lambda_\gamma)$ \cite{BLS99}
\begin{eqnarray}\label{hsEqbl}
(\partial_{\alpha\beta} -i\lambda_{\alpha}
\lambda_{\beta})\Phi(X,\lambda) = 0 \; ,
\end{eqnarray}
which may be called the ``preonic'' equation in the light of the
conjecture of \cite{BPSpreon}.

Upon the Fourier transform of $\Phi(x,\lambda)$ into
$C(X,y^\alpha)=\int\,d^n\lambda\,e^{i\lambda_\alpha\,y^\alpha}\Phi(X,\lambda)$
the equation (\ref{hsEqbl}) takes the following equivalent form
\cite{V01s}
\begin{eqnarray}\label{hsEqby}
(\partial_{\alpha\beta} + i {\partial \over \partial y^{\alpha}}
{\partial \over \partial y^{\beta}}) C(X,y) = 0 \; .
\end{eqnarray}
As was first shown in \cite{V01s} the only dynamical fields among
the components of the series expansion of
$C(X,y)=b(X)+f_\alpha(X)\, y^\alpha +
\sum^\infty_{n=2}C_{\alpha_1\cdots
\alpha_n}(X)\,y^{\alpha_1}\cdots y^{\alpha_n}$ are the scalar
field $b(X)$ and the spinor (or `svector') field $f_\alpha(X)$
which, as a consequence of (\ref{hsEqby}), satisfy the following
equations of motion
\begin{eqnarray}\label{hsEqb0}
(\partial_{\alpha\beta}\partial_{\gamma\delta}  -
\partial_{\alpha\gamma}\partial_{\beta\delta})\, b(X):=
2\partial_{\alpha[\beta}\partial_{\gamma]\delta}\, b(X) =0 \; ,
\\ \label{hsEqf0}
\partial_{\alpha\beta} f_{\gamma}(X)  -
\partial_{\alpha\gamma}f_{\beta}(X):=2\partial_{\alpha[\beta} f_{\gamma]}(X) =0 \;.
\end{eqnarray}
The equations (\ref{hsEqbl})--(\ref{hsEqf0}) are $OSp(1|2n)$
invariant \cite{V01s}, the subgroup $GL(n)$ of $OSp(1|2n)$ being a
manifest symmetry of these equations. The fields $b(X)$ and
$f_\alpha(X)$ are superpartners whose $OSp(1|2n)$ transformations
the reader can find in \cite{V01s}. Below we present only their
part which corresponds to rigid supersymmetry and superconformal
boosts with parameters $\epsilon^\alpha$ and $s_\alpha$,
respectively
\begin{eqnarray}\label{osp}
\delta
 b(X)=\epsilon^\alpha\,f_\alpha(X)+2s_\alpha\,X^{\alpha\beta}f_\beta(X)\, ,
 \quad \delta
f_\alpha(X)=\epsilon^\beta\,\partial_{\beta\alpha}\,b(X)+2s_\gamma
X^{\gamma\beta}\,\partial_{\beta\alpha}\,b(X)\, . \; \quad
\end{eqnarray}

In the case of $n=4$ (\ref{x}) the fields $b(X)$ and $f_\alpha(X)$
subject to eqs. (\ref{hsEqb0}) and (\ref{hsEqf0}) describe the
infinite tower of the massless (and thus conformally invariant)
fields of all possible integer and half--integer spins in the
physical four--dimensional subspace of the ten--dimensional
tensorial space \cite{fronsdal1,V01s}. In the cases of $n=8$ and
$n=16$ which correspond to $D=6$ and $D=10$ space--time,
respectively, the equations (\ref{hsEqb0}) and (\ref{hsEqf0})
describe conformally invariant higher spin fields with self--dual
field strengths (work in progress).

The field strengths of the $D=4$ higher spin fields are components
of the series expansion of $b(X)=b(x^l,y^{mn})$ and
$f_\alpha(X)=f_\alpha(x^l,y^{mn})$ in powers of the tensorial
coordinate $y^{mn}$
\begin{eqnarray}\label{ymn}
&~&\!\!\!\!\!\!\!\!\!\!\!\!\!\!\!\!\!
b(x^l,\,y^{mn})=\phi(x)+y^{m_1n_1}F_{m_1n_1}(x)
+y^{m_1n_1}\,y^{m_2n_2}\,[C_{m_1n_1,m_2n_2}(x)+{1\over 4}
\partial_{[n_1}\,\eta_{m_1][m_2}\,\partial_{n_2]}\phi(x)]\nonumber\\
&~&\hspace{150pt} +\sum_{s=3}^{\infty}\,y^{m_1n_1}\cdots
y^{m_sn_s}\,[C_{m_1n_1,\cdots,m_sn_s}(x)+\cdots]\,,\nonumber\\
&~&\\
 &~&
\!\!\!\!\!\!\!\!\!\!\!\!\!\!\!\!\! f^\a(x^l,y^{mn})
=\psi^\a(x)+y^{m_1n_1}[\Psi^\a_{m_1n_1}(x)+{1\over
2}\partial_{m_1}(\gamma_{n_1}\psi)^\a]\nonumber\\
&~& \hspace{150pt}+\sum_{s={5\over 2}}^{\infty}\,y^{m_1n_1}\cdots
y^{m_{s-{1\over 2}}n_{s-{1\over
2}}}\,[\Psi^\a_{m_1n_1,\cdots,m_{s-{1\over 2}}n_{s-{1\over
2}}}(x)+\cdots]\,.\nonumber
\end{eqnarray}
In (\ref{ymn}) $\phi(x)$ and $\psi^\a(x)$ are scalar and spin 1/2
field, $F_{m_1n_1}(x)$ is the Maxwell field strength,
$C_{m_1n_1,m_2n_2}(x)$ is the Weyl curvature tensor of the
linearized gravity, $\Psi^\a_{m_1n_1}(x)$ is the Rarita--Schwinger
field strength and other terms in the series stand for strengths
of spin-$s$ fields (which also contain contributions of
derivatives of lower spin fields denoted by dots, as in the case
of the Weyl curvature and of the Rarita--Schwinger field).

The $OSp(1|2n)$ invariant equations of motion of the fields $b(X)$
and $f_\alpha(X)$ propagating on the group manifold $Sp(n)$ (see
eqs. (\ref{bsp}) and (\ref{fsp}) of Section 2.2) have been derived
in \cite{Dima03} from the $Sp(n)$ counterpart of (\ref{hsEqbl})
and (\ref{hsEqby})
\begin{equation}\label{ysp}
\left[\nabla_{\a\b}-{i\over 2}(Y_\a Y_\b+Y_\b
Y_\a)\right]C(X,y)=0\,,\quad Y_\a\equiv {i} {\partial\over
{\partial y^\a}}+{\varsigma\over 4}y_\a\,,
\end{equation}
where $\nabla_{\a\b}$ are covariant derivatives generating the
algebra $Sp(n)$
\begin{equation}\label{sp}
[\nabla_{\a\b},\nabla_{\g\d}]={\varsigma }
C_{\a(\g}\nabla_{\d)\b}+{ \varsigma} C_{\b(\g}\nabla_{\d)\a},
\end{equation}
$C_{\a\b}$ is a simplectic metric and $\varsigma$ is a constant
proportional to the inverse AdS radius (the square root of the
cosmological constant).
 In \cite{Misha} the general
solution of the equations (\ref{ysp}) and their generalization to
the supergroup manifolds $OSp(N|n)$ were constructed and analyzed.
 For instance, in the case of $n=4$ the equations
(\ref{ysp}) are equivalent to an infinite system of equations of
motion of all the integer and half--integer higher spin fields
propagating in $AdS_4$ \cite{Misha,Dima03}.

At this point we should make the following comment. In the
formulation described in eqs. (\ref{hsEqbl})--(\ref{sp}) the
fields $b(X)$ and $f_\alpha(X)$ have the same statistics, namely
they are Grassmann even if $\Phi(X,\lambda)$ or $C(X,y)$ is
Grassmann even, while if we would like $b(X)$ and $f_\alpha(X)$ to
form a scalar $OSp(1|n)$ supermultiplet we should assign to
$f_\alpha(X)$ the fermionic statistics. An a priori un--physical
statistics of a part of higher spin fields is a generic feature of
the unfolded formulations of higher spin field theory involving
twistor--like Grassmann even spinor variables $\lambda_\alpha$ or
$y^\alpha$ \cite{unfold,V01s}. To single out the fields with
physically correct statistics one can use several equivalent
prescriptions \cite{mv,V99,V01s}. In our case the most appropriate
one is the following `parity conservation' requirement used in
\cite{BLS99,V01}. One should consider the complete (doubly
degenerate) spectrum of states of the quantum superparticle model
(\ref{S0}) which on the mass shell is described by a generic
Grassmann even superfield of the form
\begin{equation}\label{Upsilon1}
g_0(X,\lambda, \theta)=
\Phi(X,\lambda)+i(\lambda_\a\theta^\a)\,\Psi (X,\lambda),
\end{equation}
 where
$\Psi (X,\lambda)$ is a Grassmann odd counterpart of $\Phi
(X,\lambda)$. In $\Psi (X,\lambda)$ the half integer spin fields
have the correct statistics while the integer spin fields do not.
We now notice that the Grassmann parity of $g_0(X,\lambda,
\theta)$ can be related to the parity of $g_0(X,\lambda,\theta)$
under the change of the sign of $\lambda$ ($\lambda \rightarrow -
\lambda$). If $g_0(X,\lambda, \theta)=g_0(X,-\lambda, \theta)$
then $\Phi(X,\lambda)$ and $\Psi (X,\lambda)$ are expanded in the
integer and half integer series of $\lambda$, respectively, and
contain the fields of only physically appropriate statistics (see
\cite{BLS99} for details). Thus, strictly speaking, one should
regard the fields $b(X)$ and $f_\a(X)$ of eqs. (\ref{hsEqb0}),
(\ref{hsEqf0})  and (\ref{bsp}), (\ref{fsp}) as ones which come,
respectively, from the field $\Phi(X,\lambda)$ and $\Psi
(X,\lambda)$ of (\ref{Upsilon1}), namely
$b(X)=\int\,d^n\lambda\,\Phi(X,\lambda)$ and
$f_\a(X)=\int\,d^n\lambda\,\lambda_\a\Psi(X,\lambda)$. We shall
discuss this in more detail in Section 2.

 Since the equations (\ref{hsEqb0}), (\ref{hsEqf0})  and
their AdS counterparts are supersymmetric, a natural question
arises whether these equations can be formulated as superfield
equations in a corresponding tensorial superspace\footnote{A
superfield formulation of the unfolded higher spin dynamics
\cite{unfold,V01s} was considered in \cite{sss}.} and whether they
allow for a nonlinear generalization which would result in an
interacting theory of higher spin fields. In this paper we study
these problems.

First we combine the scalar field $b(X)$ and the spinor field
$f_\alpha(X)$ into a scalar superfield $\Phi(X,\theta)$ and find
simple superfield equations for $\Phi(X,\theta)$ which reproduce
(\ref{hsEqb0}), (\ref{hsEqf0}) and the ``preonic'' equation
(\ref{hsEqbl}). Then we look for a non--linear generalization of
the superfield equations. Our initial assumption has been that a
class of non--linear models of this kind can be constructed in a
consistent geometrical way by considering a supergravity in
tensorial superspace. A stronger conjecture might be that the
tensorial supergravity itself is an example of a theory of
interacting higher spin fields. If it was so, the
superdiffeomorphisms and the local $GL(n)$ or $SL(n)$ structure
group transformations of the tensorial superspace could generate
infinite higher spin superalgebras in ordinary space--time.

Our reasoning behind the idea to look for a non--linear dynamics
of higher spin fields within a superfield formulation of tensorial
supergravity and not, for instance within a bosonic tensorial
gravity has been two fold
\begin{itemize}
\item
 the superfield equations of motion of
the free higher spin fields are much simpler than their component
counterparts and hence may be more appropriate for a non--linear
generalization and
\item
as the experience of dealing with conventional superfield gauge
and supergravity theories teaches us, the imposition of
constraints on the superfield contents of these theories reduces
the number of possible choices and in many cases produces a
complete set of  superfield equations of motion whose form would
be otherwise hard to guess in the absence of clear
group--theoretical and geometrical guidelines.
\end{itemize}

As we shall see, the supergeometry of the tensorial supergravity
with $GL(n)$ or $SL(n)$ holonomy which we derive from the
requirement of the $\kappa$--invariance of the superparticle
action in the curved superspace background resembles that of
$N=1$, $D=2$ and $D=3$ supergravity. We find that general
solutions of the supergravity constraints are tensorial
superspaces conformally related to flat tensorial superspace or to
the supergroup manifold $OSp(1|n)$. Because of the conformal
symmetry of the supergravity constraints and of the scalar
superfield equation such a geometry is trivial in the sense that
it cannot generate a kind of `minimal' coupling of higher spin
fields to their potentials. So our expectations to find
non--linear higher spin field equations in the framework of
tensorial supergravity have not been materialized yet. However, we
believe that the results obtained lay a geometrical basis for a
new class of models formulated in tensorial superspaces and may be
useful for further development of this subject in various
directions. One of them may hopefully bring us to a non--linear
higher spin dynamics.

 The paper is organized as follows. In Section 2 we construct the equations
of motion of a scalar superfield $\Phi(X,\theta)$ in the flat
tensorial superspace and on the supergroup manifold $OSp(1|n)$. We
also find a superfield generalization of the ``preonic'' equation
(\ref{hsEqbl}) and of its AdS counterpart (\ref{ysp}).

In Section 3 we introduce the supergeometry of a curved tensorial
superspace with the holonomy group $GL(n)$ and find constraints on
its torsion and curvature which are required by the
$\kappa$--invariance of the (`preonic') superparticle action. We
then impose additional conventional supergravity constraints and
study the consistency of the torsion and curvature Bianchi
identities. In particular we find that, as in the case of $N=1$,
$D=3$ supergravity \cite{D=3}, the supergeometry with $SL(n)$
holonomy is described by an antisymmetric tensor superfield
$R_{\alpha\beta}(X,\theta)$ and by a totally symmetric field
$G_{\alpha\beta\gamma}(X,\theta)$.

Section 4 is devoted to the consideration of the dynamics of the
scalar superfield in an external tensorial supergravity
background. It is shown that its consistency requires the
background supergeometry to have $SL(n)$ holonomy.

In Section 5 we describe generalized Weyl (superconformal)
transformations of the supervielbeins and superconnection which
leave the constraints form--invariant and study superconformally
flat and $OSp$--related geometries of tensorial superspaces.

In Section 6 we show that (being superconformally invariant) the
dynamics of the scalar superfield propagating in a conformally
flat or $OSp(1|n)$--related tensorial superspace is described by
the free scalar superfield equation in flat superspace or on the
supergroup manifold $OSp(1|n)$ and hence does not lead to a
non--trivial interacting theory of higher spin fields.

The general solution of the tensorial supergravity constraints is
considered in Section 7. It is shown that (up to possible
topological subtleties) the conformal tensorial superspaces are
the only solutions of this theory.

In Conclusion we summarize the main results obtained and discuss
possible ways in which they can be developed.

\def\theequation{\arabic{section}.\arabic{equation}}
\setcounter{equation}0
\section{Superfield generalization of the massless higher spin equations}

\subsection{Scalar superfield equations in flat tensorial superspace}

Let us consider a scalar superfield
\begin{equation}\label{scalar}
\Phi (X^{\alpha\beta} ,
\theta^\gamma)=b(X)+f_\alpha(X)\,\theta^\alpha+
\sum_{i=2}^{n}\phi_{\alpha_1\cdots\alpha_i}(X)\,\theta^{\alpha_1}\cdots\theta^{\alpha_i}
\end{equation}
  in a flat
tensorial superspace whose coordinates transform under rigid
supertranslations as follows
\begin{equation}\label{flatsusy}
\delta\theta^\alpha=\epsilon^\alpha, \quad \delta
X^{\alpha\beta}={i\over 2}(\theta^\alpha\,\epsilon^\beta +
\theta^\beta\,\epsilon^\alpha)=i\theta^{(\alpha}\,\epsilon^{\beta)}\,.
\end{equation}

We are looking for a superfield equation for $ \Phi (X , \theta)$
which would reproduce the equations (\ref{hsEqb0}) and
(\ref{hsEqf0}) for the leading components of $\Phi (X , \theta)$
and from which it would follow that the higher components of the
superfield $\Phi (X , \theta)$ are completely auxiliary and vanish
on the mass shell. Since (\ref{hsEqb0}) and (\ref{hsEqf0}) are
manifestly $GL(n)$ invariant, the corresponding superfield
equation should also possess this symmetry. Taking this into
account we find that the only possible superfield equation
quadratic in supercovariant derivatives $D_\alpha= \partial
/\partial \theta^\alpha + i \theta^{\beta}
\partial_{\beta\alpha}$  $(\{D_\a,D_\b\}=2i\partial_{\a\b})$ is
\begin{eqnarray}\label{hsSEq}
D_{[\alpha}D_{\beta ]} \Phi (X, \theta ) = 0 \; .
\end{eqnarray}
It can be regarded as a generalization to the tensorial superspace
of the defining conditions of a tensor supermultiplet in $D=4$ or
of the equations of motion of a scalar supermultiplet in $D=3$.

The analysis of the equation (\ref{hsSEq}) in flat tensorial
superspace with an arbitrary even number $n$ of the Grassmann
coordinates shows that all components  of the superfield
(\ref{scalar}) subject to (\ref{hsSEq}) vanish, except for $b(X)$
and $f_\alpha(X)$, and the latter satisfy the equations
(\ref{hsEqb0}) and (\ref{hsEqf0}).

 The equation (\ref{hsSEq}) can be derived in
a rigorous way from a superfield equation which one gets by
quantizing the tensorial superparticle model (\ref{S0}). As was
considered in detail in \cite{BLS99} the quantum states of the
tensorial superparticle form a bosonic superfield
\begin{equation}\label{Upsilon}
\Upsilon(X,\theta,\lambda,\chi)=g_0(X,\theta,\lambda)+i\chi\,g_1(X,\theta,\lambda),
\end{equation}
where $\chi$ is a real single Clifford variable ($\chi^2=1$) of
the Grassmann odd parity. As has been mentioned in the
Introduction, to have the correct relation between spin and
statistics of the components of the series expansion of $g_0$ and
$g_1$ in powers of $\lambda_\alpha$, we require that
(\ref{Upsilon}) is an even function under the change of sign of
$\lambda$ and $\chi$ ($\lambda \rightarrow - \lambda,~\chi
\rightarrow - \chi$), namely
$\Upsilon(X,\theta,\lambda,\chi)=\Upsilon(X,\theta,-\lambda,-\chi)$.
This implies that $g_0(X,\theta,\lambda)=g_0(X,\theta,-\lambda)$
and $g_1(X,\theta,\lambda)=-g_1(X,\theta,-\lambda)$.

The superfield (\ref{Upsilon}) satisfies the first order
differential equation \cite{BLS99}
\begin{equation}\label{DUpsilon}
(D_\a-\chi\,\lambda_\a)\Upsilon(X,\theta,\lambda,\chi)=0.
\end{equation}

{}From (\ref{DUpsilon}) it follows that
\begin{equation}\label{01}
D_\a\,g_0-\, i\,\lambda_\a\,g_1=0, \qquad
D_\a\,g_1- \, i\,\lambda_\a\,g_0=0\,.
\end{equation}
Hence, for example $g_1$ can be expressed in terms of $D_\a\,g_0$
\begin{equation}\label{g1}
g_1= -i\mu^\alpha\,D_\a\,g_0,
\end{equation}
where $\mu^\a$ is ``inverse'' of $\lambda_\a$ in the sense that
$\mu^\a\lambda_\a=1$.

Thus only one superfield component of (\ref{Upsilon}), e.g.
$g_0(X,\theta,\lambda)=g_0(X,\theta,-\lambda)$, is independent.
Now taking the derivative $D_\a$ of (\ref{01}) we find that $g_0$
should obey the equation
\begin{equation}\label{preonsusy}
(D_{\alpha}D_{\beta }+\lambda_\a\lambda_\b)\, g_0 (X,
\theta,\lambda ) = 0 \; .
\end{equation}

The symmetric part of (\ref{preonsusy}) is
$$
(\partial_{\a\b}-i\lambda_\a\,\lambda_\b)\,g_0(X,\theta,\lambda)=0,
$$
which is similar to (\ref{hsEqbl}), while the antisymmetric part
is
\begin{eqnarray}\label{hsSEg}
D_{[\alpha}D_{\beta ]}\, g_0 (X, \theta,\lambda ) = 0 \; .
\end{eqnarray}

Thus, we can regard (\ref{preonsusy}) and/or (\ref{DUpsilon})
as a
superfield generalization of the ``preonic" equation
(\ref{hsEqbl}).

Integrating (\ref{hsSEg}) over $\lambda$ and defining $\Phi (X,
\theta)=\int\,d^n\lambda\,g_0 (X, \theta,\lambda )$, so that
$b(X)=\int\,d^n\lambda\,g_0(X,\theta,\lambda)|_{\theta=0}$ and
$f_\a(X)=\int\,d^n\lambda\,D_\a\,
g_0(X,\theta,\lambda)|_{\theta=0}$, we get the equation
(\ref{hsSEq}). Thus, on the mass shell the scalar superfield
(\ref{scalar}) is linear in $\theta^\a$, which is in accordance
with the form of the wave function describing on--shell quantum
states of the tensorial superparticle discussed in the
Introduction (eq. (\ref{Upsilon1})).

\subsection{Scalar superfield equations on OSp(1$|$n)}
Let us now consider the case when $X^{\alpha\beta}$ and
$\theta^\alpha$ parametrize a supergroup manifold $OSp(1|n)$ and
find the corresponding generalization of the superfield equation
(\ref{hsSEq}). For this we should replace the flat covariant
derivatives $D_\alpha$ with $OSp(1|n)$ covariant derivatives
$\nabla_\alpha$ which extend the $sp(n)$ algebra (\ref{sp}) to the
$osp(1|n)$ superalgebra \footnote{Explicit expressions for the
 $OSp(1|n)$ Cartan forms and
covariant derivatives in particular parametrizations has been
given in \cite{BLPS,Dima03} and for the $OSp(N|n)$ Cartan forms in
a generic parametrization in \cite{Misha}.}
\begin{equation}\label{osp1}
\{ \nabla_\alpha , \nabla_{\beta} \} = 2i \nabla_{{\alpha\beta}}
\; , \qquad [\nabla_{{ \alpha \alpha^\prime}} , \nabla_\beta ] =
{\varsigma}\, C_{\beta (\alpha } \nabla_{\alpha^\prime )}.
\end{equation}
The scalar superfield equation on the supergroup manifold
$OSp(1|n)$  has the following form
\begin{eqnarray}
\label{DD=AdS} \left(\nabla_{[\alpha} \nabla_{\beta]} +
i{\varsigma\over 4} C_{\alpha\beta} \right) \Phi(X,\theta)=0 \; .
\end{eqnarray}
The equation (\ref{DD=AdS}) reduces to the following equations  on
the dynamical components of $\Phi(X,\theta)$ \cite{Dima03}
\begin{equation}\label{bsp}
\nabla_{\a[\b}\nabla_{\g]\d}b(X)={\varsigma\over
4}\left(C_{\a[\b}\nabla_{\g]\d}+C_{\d[\b}\nabla_{\g]\a} -
C_{\b\g}\nabla_{\a\d}\right)b(X)+{\varsigma^2\over
16}\left(C_{\a\d}C_{\b\g}-C_{\a[\b}C_{\g]\d}\right)b(X),
\end{equation}
\begin{equation}\label{fsp}
\nabla_{\a[\b}f_{\g]}(X)=-{\varsigma\over
4}\left(C_{\a[\g}f_{\b]}(X)+C_{\b\g}f_\a(X)\right)\,.
\end{equation}

The coefficient in front of the second term of (\ref{DD=AdS}) is
fixed by checking the integrability of this equation. To this end
we observe that $\nabla_{\alpha [\beta} \nabla_{\gamma
]\delta}b(X)= (\nabla_{\alpha [\beta} \nabla_{\gamma
]\delta}\Phi(X,\theta))\vert_0$ and hence in view of (\ref{sp})
\begin{eqnarray}\label{compare}
\nabla_{(\alpha
[\beta} \nabla_{\gamma ]\delta )}\, b(X) &=& {1\over 2}(\nabla_{\alpha
[\beta} \nabla_{\gamma ]\delta}+\nabla_{\delta[\beta}
\nabla_{\gamma ]\alpha})b(X) = {1\over 2} [\nabla_{\alpha [\beta},
\nabla_{\gamma ]\delta}] \, b(X)
\nonumber \\
&=& {\varsigma\over 4} \left(C_{\alpha [\beta} \nabla_{\gamma
]\delta }\,b(X) + C_{\delta [\beta} \nabla_{\gamma ]\alpha
}\,b(X)- C_{\beta\gamma} \nabla_{\alpha \delta}\,b(X)\right)\,.
\end{eqnarray}
 The equation (\ref{compare})
is then compared with the bosonic equation (\ref{bsp}) which
follows from (\ref{DD=AdS}). This fixes in the latter the factor
${{i\varsigma}\over 4}$.

A superfield generalization of the ``AdS preonic'' equation
(\ref{ysp})considered in \cite{Misha} is
\begin{equation}\label{DUpsilonAds}
(\nabla_\a-\chi\,Y_\a)\Upsilon(X,\theta,\lambda,\chi)=0\,,
\end{equation}
while the $OSp(1|N)$ analog of eq. (\ref{preonsusy}) is
\begin{equation}\label{adspreonsusy}
(\nabla_{\alpha}\nabla_{\beta }+Y_\b Y_\a )\, g_0 (X,
\theta,\lambda ) = 0 \; , \qquad
Y_\a=\lambda_\a-{{i\varsigma}\over
4}\,C_{\a\b}\,{\partial\over{\partial\lambda_\b}}\,.
\end{equation}

We observe that the superfield equations (\ref{hsSEq}) and
(\ref{DD=AdS}) are much simpler than their component counterparts
and therefore it is natural to take them as a starting point in
the search for a non--linear generalization of the higher--spin
field equations. Since the scalar field contains only the
linearized field strengths of the higher spin fields one needs to
find a room for higher spin field potentials which are required
for the construction of `minimal' higher spin interactions. In
this respect on can consider supergravity in tensorial superspace
and its coupling to the scalar superfield as a model which might
provide us with minimal--like higher spin interactions via
supervielbeins and superconnections.

\def\theequation{\arabic{section}.\arabic{equation}}
\setcounter{equation}0

\section{Geometry of tensorial superspace}
\subsection{The definition of tensorial supergeometry} As in the
conventional supergravity case, curved tensorial superspace
geometry is described by the supervielbein one forms
$E^{\alpha\beta}(Z)=E^{\beta\alpha}(Z)=dZ^ME^{~~\alpha\beta}_{M}(Z)$
and $E^\alpha (Z)=dZ^ME^{~~\alpha}_{M}(Z)$. The supercoordinates
$Z^M=(X^{\mu\nu},\theta^\rho)$ are assumed to transform under the
superdiffemorphisms $Z^{\prime\, M}=f^M(Z^N)$
($\mathrm{sdet}(\partial f^M/\partial Z^N)
\not= 0$) which leave the supervielbeins
invariant ($E^{\prime \, {\cal A}}(Z^\prime) =
E^{{\cal A}}(Z)$).

We have seen that in the flat case the superparticle model
(\ref{S0}) is manifestly invariant under the rigid transformations
of the group $GL(n)$ (\ref{gl}), which can be regarded as a kind
of the ``Lorentz" group in the tensorial superspace. We shall
therefore assume that in the tensorial supergravity $GL(n)$ plays
the role of a generalized local Lorentz group acting in the
co--tangent tensorial superspace whose local basis is given by the
supervielbeins $E^{{\cal A}}=(E^{\alpha\beta},\,E^\gamma)$. As so, by
analogy with the conventional spin connection of general
relativity and supergravity we introduce the $GL(n)$ connection
\begin{eqnarray}
\label{Om} \Omega_{\beta}{}^{\alpha}:= dZ^M \Omega_{M
\beta}{}^{\alpha} \equiv E^{{\cal A}} \Omega_{{\cal
A}\beta}{}^{\alpha}\; ,
\end{eqnarray}
 the torsion 2--forms (where $\cal D$ stands for the $GL(n)$--covariant differential)
\begin{eqnarray}\label{Tb=def}
T^{\alpha\beta}& := & {\cal D} E^{\alpha\beta} \equiv
dE^{\alpha\beta} - E^{\alpha\gamma}\wedge \Omega_{\gamma}{}^{\beta
}-E^{\beta\gamma}\wedge
\Omega_{\gamma}{}^{\alpha }\; , \\
\label{Tf=def} T^{\alpha}& := & {\cal D} E^{\alpha} \equiv
dE^{\alpha} - E^{\beta}\wedge \Omega_{\beta}{}^{\alpha} \; ,
\end{eqnarray}
and the curvature of the $GL(n)$ connection
\begin{eqnarray}
 \label{R=def}
{\cal R}_{\beta}{}^{\alpha}:= d\,
 \Omega_{\beta}{}^{\alpha} -
\Omega_{\beta}{}^{\gamma}\wedge \Omega_{\gamma}{}^{\alpha} \; .
\end{eqnarray}
 The Ricci identity ${\cal D}{\cal D} = {\cal R}$ in our
 notation implies ${\cal D}{\cal D}E^{\alpha}\equiv - E^{\beta}\wedge  {\cal
R}_{\beta}{}^{\alpha}$.

In what follows we shall also discuss consequences of the
restriction of the $GL(n)$ curvature to the $SL(n)$ curvature by
imposing the tracelessness constraint ${\cal
R}^{~\alpha}_\alpha=0$.

The next step is to find the constraints on tensorial
supergeometry. In the case of conventional super Yang--Mills and
supergravity theories there are different geometrical and physical
guiding lines to get superfield constraints. The one which we have
at our disposal is the $\kappa$--symmetry of the massless
superparticle.

\subsection{The massless superparticle in curved tensorial superspace.}
Let us consider the dynamics of a superparticle in a curved
tensorial superspace and find restrictions on its supergeometry
which follow from the requirement for the model to possess the
same symmetries as in the flat limit. Thus, we shall derive the
constraints on torsion and curvature of a supergravity in
tensorial superspace using the conventional requirement that a
superparticle or a superbrane propagating in the supergravity
background should be invariant under $\kappa$--symmetry, as in the
flat case.

\subsubsection{Superparticle action, $\kappa$--symmetry and the basic torsion constraint
in tensorial  superspace}

A straightforward generalization of the action (\ref{S0}) to the
curved tensorial superspace is
\begin{eqnarray}\label{S}
S
= {1\over 2} \int {E}^{\alpha\beta}(Z(\tau))\,
\lambda_{\alpha}(\tau)\, \lambda_{\beta} (\tau)\;
= {1\over 2} \int d\tau {E}_\tau^{\alpha\beta}
\lambda_{\alpha}\lambda_{\beta}
\; ,
\end{eqnarray}
where the flat superform $dX^{\a\b}(\tau) -id\theta^{(\a}\theta^{\beta)}(\tau)$
of eq. (\ref{S0}) has been replaced with the pull--back on the
superparticle worldline of the bosonic supervielbein form
${E}^{\alpha\beta}(Z)$
\begin{eqnarray}
\label{hEab} {E}^{\alpha\beta}(Z(\tau)) := d\tau
{E}_\tau^{\alpha\beta}=
 d{Z}^{M}(\tau) E_M^{\alpha\beta}({Z}(\tau))\;.
\end{eqnarray}

In the flat case the action (\ref{S0}) is invariant under local
$\kappa$--symmetry with $n-1$ independent parameters, which means
that the superparticle under consideration can be associated with
a BPS state (called the BPS preon \cite{BPSpreon}) which preserves
all but one supersymmetry \cite{BL98}. The $\kappa$--symmetry
transformations of the action (\ref{S0}) are
\begin{eqnarray}\label{kap0b}
\delta_{\kappa} {X}^{\alpha\beta}(\tau) = i \delta_{\kappa}
{\theta}^{(\alpha}(\tau)\, {\theta}^{\beta )}(\tau) \; , \quad
\delta_{\kappa}\, {\lambda}_{\alpha}(\tau)=0\; , \quad
\end{eqnarray}
\begin{equation}\label{kap0f0}
\delta_{\kappa} {\theta}^{\alpha}(\tau) = \sum_{I=1}^{n-1}
\kappa^I (\tau) \,\mu^{\alpha}_I(\tau) \; ,
\end{equation}
where $\kappa^I(\tau)$ are $n-1$ fermionic parameters and
$\mu^{\alpha}_I(\tau)$ is a set of $(n-1)$ auxiliary {\it bosonic}
$GL(n)$ vectors (or spinors of an $SO(t,D-t)\subset GL(n)$ for
$n=2^k$) which are orthogonal to $\lambda_{\alpha}(\tau)$
\footnote{The bosonic spinors $\mu^\alpha_I$ can be considered
\cite{csg} as counterparts of the Killing spinors corresponding to
an $(n-1)/n$ supersymmetric (BPS preon) solution of supergravity
equations, which is still hypothetical for the standard
supergravity but which exists in a Chern--Simons like supergravity
\cite{csg}.},
\begin{eqnarray}\label{uIl}
 \mu^{\alpha}_I(\tau) \lambda_\alpha (\tau) =0 \; , \quad I=1,\ldots (n-1)\; .
\end{eqnarray}

Actually (\ref{kap0f0}) describes the general solution of the
equation
\begin{eqnarray}\label{kap0f}
\delta_{\kappa}{\theta}^{\alpha}(\tau) \lambda_\alpha (\tau) =0 \; ,
\end{eqnarray}
which can be used instead of (\ref{kap0f0}) as the definition of
$(n-1)$ parametric $\kappa$--symmetry.

The flat superspace action (\ref{S0}) is also invariant  under
 the $n(n-1)/2$ parametric bosonic $b$--symmetry \cite{BL98,BLS99},
 which can be treated as a bosonic `superpartner' of the $\kappa$--symmetry,
\begin{eqnarray}\label{b-sym0}
\delta_b X^{\alpha\alpha^\prime} = \mu^{\alpha}_I
\mu^{\alpha^\prime}_J b^{IJ} (\tau)\;  (\Leftrightarrow  \;
\delta_b X^{\alpha\alpha^\prime} \,\lambda_{\alpha^\prime}=0 ) \;
, \qquad \delta_b \theta^{\alpha}(\tau)=0 \; , \qquad \delta_b
\lambda_{\alpha}(\tau)=0 \; .
\end{eqnarray}

 We would like the $\kappa$--symmetry as well as the
$b$--symmetry to be also preserved in the supergravity background
(see \cite{kappa}). Such a requirement has a deep physical
meaning: it implies that the limit of flat superspace (when the
background fields tends to zero) is smooth and, in particular,
that the number of the degrees of freedom of the dynamical system
does not change in such a limit. The curved superspace
generalization of the $\kappa$--symmetry and of the $b$--symmetry
transformations (\ref{kap0b}) and (\ref{b-sym0}) of the coordinate
functions are, respectively,
\begin{eqnarray}\label{kappaGb1}
i_\kappa E^{{\alpha \alpha^\prime}}:= \delta_\kappa Z^{{M}} E_{
{M}}^{{\alpha \alpha^\prime}} =0 \; , \quad
 i_\kappa E^{\alpha} :=  \delta_\kappa Z^{{M}}
E_{ {M}}^{\alpha} = \mu^{\alpha}_I\kappa^I(\tau)\; . \quad
\end{eqnarray}
and
\begin{eqnarray}\label{b-sym}
i_b E^{\alpha\alpha^\prime} := \delta_b Z^{{M}} E_{ {M}}^{{\alpha
\alpha^\prime}} = \mu^{\alpha}_I \mu^{\alpha^\prime}_J b^{IJ}
(\tau)\;  ,   \quad
 i_b E^{\alpha} :=  \delta_b Z^{{M}}
E_{ {M}}^{\alpha} =0 \; .
\end{eqnarray}
The variation of the bosonic spinor field
$\lambda_{\alpha}(\tau)$, $\delta_\kappa \lambda_{\alpha}$ and
$\delta_b \lambda_{\alpha}$ are to be defined from the invariance
of the action.

The invariance of the action (\ref{S}) under the $\kappa$-- and
$b$--transformations (\ref{kappaGb1}) and (\ref{b-sym}) requires
the bosonic torsion of the tensorial superspace to be restricted
by the constraints\footnote{We should note that the requirement of
the $\kappa$--symmetry itself already leads to the constraints
(\ref{Tb=V2}), while taking into the consideration of the
$b$--invariance makes the analysis simpler.}
\begin{eqnarray}\label{Tb=V2}
T^{\alpha\beta}&  =-i E^{\alpha}\wedge E^{\beta} +
2E^{\gamma}\wedge E^{\delta(\alpha}  t_{\gamma
\delta}{}^{\beta)}(Z)+
  E^{\gamma\gamma^\prime}\wedge E^{\delta  (\alpha} \;
 t_{\gamma\gamma^\prime\;  \delta }{}^{\beta )}(Z) \; .
\end{eqnarray}
The complete set of the $\kappa$--symmetry and $b$--symmetyr
transformations leaving the action  (\ref{S}) invariant in the
background (\ref{Tb=V2}) is
\begin{eqnarray}\label{kappaGb2}
i_\kappa E^{{\alpha \alpha^\prime}}=0  \; , \quad
 i_\kappa E^{\alpha} \lambda_\alpha =0 \; (\Leftrightarrow \;
i_\kappa E^{\alpha} = \mu^{\alpha}_I\kappa^I(\tau)\; ) \; , \quad
\nonumber \\  \delta_\kappa \lambda_{\alpha} = i_\kappa E^{\beta}
t_{\beta\; \alpha}{}^\gamma \lambda_\gamma \; ;
\end{eqnarray}
\begin{eqnarray}\label{b-symm}
i_b E^{{\alpha \beta}}\lambda_\beta =0 \; (\Leftrightarrow \; i_b
E^{\alpha\alpha^\prime} = \mu^{\alpha}_I \mu^{\alpha^\prime}_J
b^{IJ} (\tau)\; ) \; ,   \quad
 i_b E^{\alpha} =0 \; , \quad \nonumber \\
\delta_b \lambda_{\alpha} = {1\over 2}  i_b E^{\beta\beta^\prime}
t_{\beta\beta^\prime\; \alpha}{}^\gamma \lambda_\gamma \; . \quad
\end{eqnarray}

Eq. (\ref{Tb=V2}) is the starting point for our analysis of
possible supergravity constraints in the tensorial superspace. In
addition to (\ref{Tb=V2}) we  also impose conventional constraints
which express some of superfields in terms of other ones or,
equivalently,  fix an arbitrariness in the definition of the
supervielbeins and the connection.

As this point, although well known in the context of standard
supergravity  \cite{1001}, is important for understanding that the
supergravity constraints we find are indeed the most general ones
for the superspaces with {\it the $GL(n)$ and $SL(n)$ structure
group}, we are going to discuss it in more detail. The reader who
is not interested in technicalities may skip the next Subsection
3.2.2 and pass directly to Subsection 3.3.

\subsubsection{On the freedom in superfield redefinitions and conventional
constraints}

In the case under consideration it is essential that the $GL(n)$
structure group symmetry of the tensorial superspace allows one to
make, for instance, the following redefinition of the `spin'
connection (\ref{Om}) and of the fermionic supervielbein
\begin{eqnarray}\label{Omred}
\Omega_{\alpha}{}^{\beta} \; \mapsto  \; \Omega_{\alpha}{}^{\beta}
+ E^\gamma r_{\gamma \alpha}{}^{\beta} + {i\over 2}
E^{\gamma\delta} r_{\gamma \delta}{}_{\alpha}{}^{\beta} \; ,
\end{eqnarray}
\begin{equation}\label{fered}
E^{\alpha} \; \mapsto  \; E^{\alpha} - E^{\beta\gamma}
S_{\gamma\beta}^{\alpha}\,,
\end{equation}
with arbitrary superfields $r_{\gamma \alpha}{}^{\beta}(Z)$,
$r_{\gamma  \delta}{}_{\alpha}{}^{\beta}(Z)= r_{\delta
\gamma}{}_{\alpha}{}^{\beta}(Z)$ and
$S_{\gamma\beta}^{\alpha}(Z)$.

We now notice that the tensorial structure of components
(\ref{Tb=V2}) of the bosonic torsion (\ref{Tb=def}) is similar to
that of the superfields which are used in the redefinitions
(\ref{Omred}) and (\ref{fered}). This allows us to simplify the
torsion (\ref{Tb=V2}) by removing the $t_{\gamma
\delta}{}^{\beta}(Z)$ superfield  and also set to zero either the
lowest dimensional component of the $GL(n)$ curvature,
$R_{\delta\; \gamma}{}_{\alpha}{}^{\beta}=0$,  or  alternatively
to eliminate the highest dimensional component of the bosonic
torsion, $t_{\gamma\gamma^\prime\;  \delta }{}^{\beta}(Z)=0$. The
additional conditions on the torsion and/or curvature obtained in
this way are called \cite{SG80} {\it conventional} constraints in
contrast to the {\it essential} constraint on the torsion given by
the form of the first term ($-i\,E^\alpha\wedge E^\beta$) on the
right hand side of (\ref{Tb=V2}).

Thus the two natural choices of the conventional constraints are
\begin{eqnarray}\label{Tb=V3}
T^{\alpha\beta}& =& -i E^{\alpha}\wedge E^{\beta} +
 E^{\gamma\gamma^\prime}\wedge E^{\delta  (\alpha} \;
 t_{\gamma\gamma^\prime\;  \delta }{}^{\beta )}(Z) \; ,
\\
 \label{R=V3}
{\cal R}_{\beta}{}^{\alpha} & = & E^{\gamma\gamma^\prime}\wedge
E^{\delta} {\cal R}_{{\delta} \;
\gamma\gamma^\prime}{}_{\beta}{}^{\alpha} + {1\over 2}
E^{\gamma\gamma^\prime}\wedge E^{\delta \delta^\prime} {\cal
R}_{{\delta \delta^\prime}\; \gamma\gamma^\prime}{}
_{\beta}{}^{\alpha} \;
\end{eqnarray}
and
\begin{eqnarray}\label{Tb=V4}
T^{\alpha\beta}& =  & -i E^{\alpha}\wedge E^{\beta}  \; ,
\\
 \label{R=V4}
{\cal R}_{\beta}{}^{\alpha} & = & {1\over 2}
 E^{\gamma}\wedge E^{\delta} \;
 {\cal R}_{\gamma\delta\;  \beta }{}^{\alpha}(Z) +
E^{\gamma\gamma^\prime}\wedge E^{\delta} {\cal R}_{{\delta} \;
\gamma\gamma^\prime}{}_{\beta}{}^{\alpha} + {1\over 2}
E^{\gamma\gamma^\prime}\wedge E^{\delta \delta^\prime} {\cal
R}_{{\delta \delta^\prime}\; \gamma\gamma^\prime}{}
_{\beta}{}^{\alpha} \; .
\end{eqnarray}
One can see that the constraints  (\ref{Tb=V3}), (\ref{R=V3}) and
(\ref{Tb=V4}), (\ref{R=V4}) are related by the redefinition
$\Omega_{\alpha}{}^{\beta} \; \mapsto  \;
\Omega_{\alpha}{}^{\beta} + {1\over 2} E^{\gamma\delta}
t_{\gamma\; \delta}{}_{\alpha}{}^{\beta}$ and  ${\cal R}_{\gamma\;
\delta}{}_{\alpha}{}^{\beta}= - it_{\gamma\;
\delta}{}_{\alpha}{}^{\beta}$.

 The consistency of the constraints (\ref{Tb=V3}), (\ref{R=V3}) or
 (\ref{Tb=V4}), (\ref{R=V4}) should be studied with the use of the
 Bianchi identities
\begin{eqnarray}\label{BIsTb}
{\cal D}T^{\alpha\beta} +  E^{\alpha\gamma}\wedge {\cal
R}_{\gamma}{}^{\beta} + E^{\beta\gamma}\wedge {\cal
R}_{\gamma}{}^{\alpha} & \equiv & 0 \; , \\ \label{BIsTf} {\cal
D}T^{\alpha} + E^{\beta}\wedge {\cal R}_{\beta}{}^{\alpha} &
\equiv & 0 \; ,
\\ \label{BIsR}  {\cal D}{\cal R}_{\beta}{}^{\alpha}  & \equiv & 0
\; .
\end{eqnarray}
It is well known (see \cite{1001}) that although in the absence of
constraints the Bianchi identities only imply that the torsion and
curvature are constructed from the supervielbeins and connection,
when the set of essential and conventional constraints are
imposed, the Bianchi identities lead to additional restrictions on
the form of the torsion and curvature, and in some cases produce
dynamical equations of motion which then imply that corresponding
supergravity constraints are on shell.

Also in our case the Bianchi identities impose further conditions
on the form of torsion and curvature. In particular, already the
study of the simplest lower dimensional component of the Bianchi
identity (\ref{BIsTb}) shows that (\ref{Tb=V3}), (\ref{R=V3}) (as
well as (\ref{Tb=V4}), (\ref{R=V4})) imply that
$T_{\gamma\beta}{}^{\alpha}=0$, {\it i.e.} that
\begin{eqnarray}\label{Tf=V3}
T^{\alpha}& =  & E^{\gamma\gamma^\prime}\wedge E^{\delta}
T_{{\delta} \; \gamma\gamma^\prime}{}^{\alpha} + {1\over 2}
E^{\gamma\gamma^\prime}\wedge E^{\delta \delta^\prime} T_{{\delta
\delta^\prime}\; \gamma\gamma^\prime}{}^{\alpha} \; .
\end{eqnarray}
Moreover, the higher dimensional components of the Bianchi
identity (\ref{BIsTb}) imply that all the superfields in $T^A$ and
$R^{\; \alpha}_\beta$ can be expressed in terms of an
antisymmetric superfield ${\cal R}_{\alpha\beta}$, a superfield
$U_{\alpha\beta\; \gamma}=U_{\beta\alpha\; \gamma}$ and their
derivatives, as we shall see in the next subsection.

\subsection{The Bianchi identities and the complete set of the constraints
in the tensorial superspace with the structure group $GL(n)$}

Thus  eq. (\ref{Tb=V2}) which follows from the requirement of the
$\kappa$--symmetry of the tensorial  superparticle  and contains
what is usually called {\it essential constraints} (in
conventional superspace these are $T_{\alpha\beta}{}^{a}= - 2i
\Gamma_{\alpha\beta}^a$) is the starting point for our analysis of
the  supergravity constraints in tensorial superspace with the
$GL(n)$ structure group. In addition to (\ref{Tb=V2}) we also
impose conventional constraints (see Subsecion 3.2.2) which
express some of superfields in terms of other ones or,
equivalently,  fix an arbitrariness in the definition of the
supervielbeins and of the connection.  By imposing the
conventional constraints and studying the Bianchi identities
(\ref{BIsTb}), (\ref{BIsTf}) and (\ref{BIsR}) one finds the form
of the torsion and curvature of the tensorial superspace.
 A particular choice of conventional constraints (see eqs.
 (\ref{Tb=V3}) and (\ref{R=V3})) results in
\begin{eqnarray}\label{Tb=AdS}
T^{\alpha\beta}& = & -i E^{\alpha}\wedge E^{\beta} + 2
E^{\gamma(\alpha}\wedge E^{\beta)\delta } R_{\gamma\delta }(Z) \;,
\\ \label{Tf=AdS} T^{\alpha}& = & 2 E^{\alpha\beta}\wedge
E^{\gamma} R_{\beta\gamma } + E^{\alpha\beta}\wedge
E^{\gamma\delta } U_{\beta{\gamma\delta }}
\; , \\
\label{R=AdS} {\cal R}_\beta{}^\alpha &=& i E^{\gamma \delta}
\wedge E^\alpha U_{\beta{\gamma\delta }} - E^{\alpha \gamma}
\wedge E^\delta (F_{\delta {\beta\gamma}}+ {\cal D}_\delta
R_{\beta\gamma}) - \nonumber \\ &-& E^{\alpha \gamma} \wedge
E^{\delta\epsilon} ({\cal D}_{(\beta} U_{\gamma )
{\delta\epsilon}} + {\cal D}_{\delta\epsilon} R_{\beta\gamma}) \;.
\end{eqnarray}
In (\ref{R=AdS}) $R_{\gamma\delta }(Z)$ and $U_{\alpha {\beta
\gamma}}(Z)= U_{\alpha {\gamma\beta}}(Z)$
 are
`main' superfields \footnote{We hope that the reader will not
confuse the curvature two--form
 ${\cal R}_\beta^{~\alpha}$  with the superfield
$R_{\alpha\beta}$. The notation for the latter has been chosen by
analogy with $N=1$, $D=3$ supergravity where
$R_{\alpha\beta}=\epsilon_{\alpha\beta}\,R$ \cite{D=3}. Note also
that, since we deal with the holonomy groups $GL(n)$ and $SL(n)$,
for $n>2$
there is no metric to rise and lower the indices.}, and
\begin{eqnarray}\label{F123=}
F_{\alpha {\beta\gamma}}= 2i U_{(\beta {\gamma )\alpha}} -
iU_{\alpha {\beta\gamma}} - 2 {\cal D}_{(\beta} R_{\gamma )
\alpha}\;.
\end{eqnarray}

The main superfields are related by the equations
\begin{eqnarray}\label{DU=DR}
{\cal D}_{[\alpha}U_{\beta] {\gamma\delta}} = - {\cal
D}_{\gamma\delta} R_{\alpha\beta} \;,
\end{eqnarray}
\begin{eqnarray}\label{DU=DF}
{\cal D}_{(\alpha}U_{\beta)\, {\gamma\delta}} = - i {\cal
D}_{(\gamma} F_{\delta )\; \alpha\beta} \;
\end{eqnarray}
and
\begin{equation}\label{5/2}
{\cal D}_{\a \b} U_{\g\delta \sigma} - {\cal D}_{\delta \sigma}
U_{\gamma \a \b} +2  U_{\g \a(\sigma} R_{\delta)\b} +
2U_{\g\b(\sigma} R_{\delta)\a} = 0\,,
\end{equation}
which are the constraints on $R_{\a\b}$ and $U_{\a\b\g}$ required
by the Bianchi identities (\ref{BIsTb}) and (\ref{BIsTf}). Due to
a straightforward generalization of the Dragon theorem
\cite{dragon} no other independent constraints arise from the
curvature Bianchi identities (\ref{BIsR}).

The superfields $U_{\alpha\beta\gamma}$ and
$F_{\alpha\beta\gamma}$ can be alternatively expressed in terms of
a totally symmetric superfield $G_{\alpha\beta\gamma}$, a
derivative of $R_{\gamma\alpha}$ and a mixed symmetry superfield
$H_{\alpha\beta\gamma}=H_{\alpha\gamma\beta}$   as follows
\begin{eqnarray}\label{UF}
U_{\alpha\beta\gamma}&=&G_{{\alpha\beta \gamma }} + {2i\over 3}
{\cal D}_{(\beta} R_{\gamma )\alpha} +
H_{\alpha\beta\gamma}\,,\nonumber\\
-iF_{\alpha\beta\gamma}&=&G_{{\alpha\beta \gamma }} + {2i\over 3}
{\cal D}_{(\beta} R_{\gamma )\alpha} -2H_{\a\b\g}\,.
\end{eqnarray}
This decomposition is useful when we perform the reduction of
$GL(n)$ holonomy to $SL(n)$--holonomy, which is achieved  by
putting $H_{\a\b\g}=0$ (see Section 4).

Note that if we choose $R_{\alpha\beta}=-{\varsigma\over 2}
C_{\alpha\beta}$ and $U_{(\alpha {\beta \gamma})}(Z)=0$ we find
that ${\cal R}_\alpha^{~\beta}=0$, and the constraints
(\ref{Tb=AdS})--(\ref{R=AdS}) reduce to the defining relations of
the Maurer--Cartan forms and of the torsion of the supergroup
$OSp(1|n)$ in the flat basis $(\Omega_\alpha^{~\beta}=0)$, whose
covariant derivatives form the $OSp(1|n)$ superalgebra (\ref{sp}),
(\ref{osp1}). The $OSp(1|n)$ Maurer--Cartan equations are
\begin{eqnarray}\label{OSPMC}
d {\cal E}^{\alpha\beta} & = &- i {\cal E}^{\alpha} \wedge {\cal
E}^{\beta} - \zeta {\cal E}^{\alpha\gamma} \wedge
{\cal E}^{\delta\beta} C_{\gamma\delta}  \; , \nonumber \\
d {\cal E}^{\alpha}& = &  - \zeta  {\cal E}^{\alpha\gamma} \wedge
{\cal E}^{\delta} C_{\gamma\delta}  \; .
\end{eqnarray}

A different but equivalent set of constraints can be obtained by
making the following redefinition of the connection
\begin{eqnarray}\label{Om-Om}
\Omega_\beta{}^\alpha \longrightarrow \Omega_\beta{}^\alpha -
E^{\alpha\gamma} R_{\gamma\beta}\;
\end{eqnarray}
which results in the corresponding redefinition of the $\sl
vector$ covariant derivative. The constraints take the form
\begin{eqnarray}\label{Tb=M}
T^{\alpha\beta}& = & -i E^{\alpha}\wedge E^{\beta}  \; , \\
\label{Tf=M}
T^{\alpha} & = & \;\; E^{\alpha\beta}\wedge E^{\gamma}
R_{\beta\gamma} + E^{\alpha\beta}\wedge E^{\gamma\delta } U_{\beta
{\gamma\delta }}\; , \\
\label{R=M} {\cal R}_\beta{}^\alpha & = & - i E^\alpha \wedge
E^{\gamma} R_{\beta\gamma} + i E^{\gamma \delta} \wedge E^\alpha
U_{\beta{\gamma\delta }}
 - E^{\alpha \gamma}
\wedge E^\delta F_{\delta {\beta\gamma}} - \nonumber \\
&& \qquad {} \qquad  - E^{\alpha \gamma}
\wedge E^{\delta\epsilon} ({\cal D}_{(\beta} U_{\gamma )
{\delta\epsilon}} + R_{\beta\epsilon}\, R_{\gamma\delta})\;,
\end{eqnarray}
where the main superfields $U_{\alpha\beta\gamma}$ and
$R_{\alpha\beta}$ satisfy the constraints
\begin{eqnarray}\label{DU=DR1} {\cal
D}_{[\alpha}U_{\beta] {\gamma\delta}} &=& - {\cal D}_{\gamma\delta}
R_{\alpha\beta} \; ,
\\
\label{DU=DF1}
{\cal D}_{(\alpha}U_{\beta)\, {\gamma\delta}} &=& - i {\cal
D}_{(\gamma} F_{\delta )\; \alpha\beta} \;
\end{eqnarray}
and
\begin{equation}\label{5/21}
{\cal D}_{\a \b} U_{\g\delta \sigma} - {\cal D}_{\delta \sigma}
U_{\gamma \a \b} + R_{\gamma (\a} U_{\b) \delta\sigma} - R_{\gamma
(\delta} U_{\sigma)\a \b} = 0\,.
\end{equation}

 To conclude, eqs. (\ref{Tb=AdS}), (\ref{Tf=AdS})
and (\ref{R=AdS}) describe the most general constraints on the
geometry of curved tensorial superspace with the holonomy group
$GL(n)$ which are required by the tensorial superparticle with
$(n-1)$ $\kappa$--symmetries. The equivalent set of constraints
(\ref{Tb=M}), (\ref{Tf=M}) and (\ref{R=M}) can be obtained by
making superfield redefinitions with the use of the main
superfields $R$ and $U$ as parameter functions.

\subsection{Tensorial superspace with the holonomy group $SL(n)$}
When $n=2$ the constraints (\ref{Tb=AdS})--(\ref{R=AdS}) and
(\ref{Tb=M})--(\ref{R=M}) describe conformal $N=1$, $D=3$
supergravity \cite{D=3}. In this case the superfield
$R_{\alpha\beta}$ gets reduced to the scalar density $R$
($R_{\alpha\beta}=\epsilon_{\alpha\beta}\, R$), and the trace part
of the $GL(2)$ connection and curvature correspond to local Weyl
(scaling) symmetry. To reduce the conformal $D=3$ supergravity to
the off--shell $N=1$, $D=3$ Poincare supergravity one imposes
additional tracelessness constraint on the curvature
\begin{eqnarray}\label{Raa=0}
{\cal R}_\alpha{}^\alpha=0\;.
\end{eqnarray}
This reduces $GL(2)$ down to $SL(2)\approx O(1,2)$ which is
isomorphic to the $D=3$ Lorentz group.

The constraint (\ref{Raa=0}), restricting $GL(n)$ to $SL(n)$, can
also be imposed in the general case of $n\geq 2$. Then the main
superfields reduce to
\begin{eqnarray}\label{U=W+DR}
 -iF_{\alpha {\beta\gamma }}=U_{\alpha {\beta\gamma }}= G_{{\alpha\beta \gamma }}
+ {2i\over 3}  {\cal D}_{(\beta} R_{\gamma )\alpha} \; ,
\end{eqnarray}
where $G_{{\alpha\beta \gamma }}$ is totally symmetric. In view of
(\ref{UF}) we observe that the condition of $SL(n)$ holonomy
amounts to putting to zero the tensor $H_{\alpha\beta\gamma}$.

The superfields $U$, $G$ and $R$ satisfy the following
differential relations
\begin{eqnarray}
\label{DU=DU} {\cal D}_{(\alpha}U_{\beta) {\gamma\delta }} ={\cal
D}_{(\gamma} U_{\delta ){\alpha\beta}}
 \; , \qquad
\end{eqnarray}
(which is a consequence of (\ref{DU=DF}) and (\ref{U=W+DR})), and
\begin{eqnarray}
\label{DW=DR} {\cal D}_{[\alpha}  G_{{\beta ] \gamma\delta }} = -
{\cal D}_{\gamma\delta} R_{\alpha\beta} - {i\over 3} ({\cal
D}_{\alpha }{\cal D}_{(\gamma} R_{\delta) \beta}-{\cal D}_{\beta
}{\cal D}_{(\gamma} R_{\delta)\alpha})\; .
\end{eqnarray}
Since $G_{\beta\gamma\delta}$ is totally symmetric, from eq.
(\ref{DW=DR}) we can get
\begin{equation}\label{DG=DR}
2{\cal D}_{[\alpha}  G_{{\beta ] \gamma\delta }} = - {\cal
D}_{(\gamma\delta} R_{\alpha)\beta} + {\cal D}_{(\gamma\delta}
R_{\beta)\alpha}\,.
\end{equation}
To derive (\ref{DG=DR}) we first symmetrize the left and the right
hand side of (\ref{DW=DR}) in $(\gamma\delta\alpha)$ and then sum
up the result with the symmetrization of (\ref{DW=DR}) in
$(\gamma\delta\beta)$.

Comparing (\ref{DW=DR}) with (\ref{DG=DR}) we find a condition
which must be satisfied by $R_{\alpha\beta}$ and which will appear
in Section 4 as part of the integrability of a scalar superfield
equation in an external tensorial supergravity background. This
condition can also be obtained by antisymmetrizing the indices
$[\alpha\beta\gamma]$ in (\ref{DW=DR}) which gives
\begin{eqnarray}\label{DDR+}
{\cal D}_{[\alpha }{\cal D}_{\beta} R_{\gamma]\delta} + {\cal
D}_{\delta} {\cal D}_{[\alpha }R_{\beta\gamma]} = 5i {\cal
D}_{\delta[\alpha }R_{\beta \gamma]} \;.
\end{eqnarray}
Then symmetrizing eq. (\ref{DDR+}) with respect to
$(\gamma\delta)$ and regrouping indices we get
\begin{eqnarray}\label{compare3}
{\cal D}_{\gamma }  {\cal D}_{[\alpha } R_{\beta] \delta} + {\cal
D}_{\delta }  {\cal D}_{[\alpha } R_{\beta] \gamma} = 2 {i}{\cal
D}_{\gamma\delta } R_{\alpha\beta} + 3i {\cal D}_{\gamma [\alpha }
R_{\beta]\delta} + 3i {\cal D}_{\delta [\alpha } R_{\beta]\gamma}
\; .
\end{eqnarray}
We shall encounter this last form of the condition on
$R_{\alpha\beta}$  in Section 7 analyzing the consistency of the
propagation of a scalar field in a non--linear tensorial
supergravity background.

Let us also note that using the anticommutation relation $\{{\cal
D}_\alpha,{\cal D}_\alpha\}=2i {\cal D}_{\alpha\beta}$, from
(\ref{compare3}) one finds that the last two terms in the right
hand side of (\ref{DW=DR}) can be rewritten in the form
\begin{equation}\label{last2}
{\cal D}_{\alpha }{\cal D}_{(\gamma} R_{\delta) \beta}-{\cal
D}_{\beta }{\cal D}_{(\gamma} R_{\delta)\alpha}=2i{\cal
D}_{\gamma\delta} R_{\alpha\beta}+i{\cal D}_{\gamma[\alpha}
R_{\beta]\delta} +i{\cal D}_{\delta[\alpha} R_{\beta]\gamma}\,
\end{equation}
which upon the substitution into (\ref{DW=DR}) gives
(\ref{DG=DR}). This can be regarded as a check or as an
alternative derivation of eq. (\ref{DG=DR}).

 Eq. (\ref{DDR+}) is identically satisfied in the
case of $N=1$, $D=3$ supergravity (in which case $\alpha, \beta
,\gamma =1,2$, and hence the antisymmetrization of three indices
gives zero), but it is nontrivial for the tensorial superspaces
with $n>2$.

Using (\ref{U=W+DR}), from eq. (\ref{DU=DU}) one derives another
consequence of the constraint (\ref{Raa=0})
\begin{eqnarray}\label{DWDW=DR}
{\cal D}_{[\alpha}  G_{{\beta] \gamma}\delta}+{\cal D}_{[\gamma}
G_{{\delta] \alpha \beta }} = - {2\over 3}\left( {\cal D}_{\delta
(\alpha} R_{\gamma)\beta }+{\cal D}_{\beta (\alpha}
R_{\gamma)\delta }\right) \; .
\end{eqnarray}
In view of (\ref{DG=DR}) the equation (\ref{DWDW=DR})  (and hence
(\ref{DU=DU})) is identically satisfied and therefore does not put
further restrictions on the form of $R_{\alpha\beta}$ and
$G_{\alpha\beta\gamma}$.

To conclude, when the holonomy group is restricted to $SL(n)$ by
(\ref{Raa=0}), the constraints (\ref{Tb=AdS}) and (\ref{Tf=AdS})
remain the same
\begin{eqnarray}
T^{\alpha\beta}& = & -i E^{\alpha}\wedge E^{\beta} + 2
E^{\gamma(\alpha}\wedge E^{\beta)\delta } R_{\gamma\delta } \;,
\nonumber \\ \label{Tsl1} T^{\alpha}& = & 2 E^{\alpha\beta}\wedge
E^{\gamma} R_{\beta\gamma } + E^{\alpha\beta}\wedge
E^{\gamma\delta } U_{\beta{\gamma\delta }} \;
\end{eqnarray}
while (\ref{R=AdS}) reduce to
\begin{eqnarray}\label{Tsl2}{\cal
R}_\beta{}^\alpha &=& i E^{\gamma \delta} \wedge E^\alpha
U_{\beta{\gamma\delta }} - E^{\alpha \gamma} \wedge E^\delta
(iU_{\delta {\beta\gamma}}+ {\cal D}_\delta R_{\beta\gamma}) -
\nonumber \\ &-& E^{\alpha \gamma} \wedge E^{\delta\epsilon}
({\cal D}_{(\beta} U_{\gamma ) {\delta\epsilon}} + {\cal
D}_{\delta\epsilon} R_{\beta\gamma}) \;.
\end{eqnarray}
The superfield $U_{\a\b\g}$ is expressed through the totally
symmetric superfield $G_{\a\b\g}$ and a derivative of the
superfield $R_{\a\b}$ by the equation (\ref{U=W+DR}), the main
superfields $G_{\a\b\g}$ and $R_{\a\b}$ being related to and
constrained by eqs. (\ref{DW=DR}) and (\ref{5/2}).

We should note that further reduction of the $SL(n)$ holonomy
group down to its subgroup $Sp(n)$ imposes in the case of $n>2$
additional restrictions on $R_{\alpha\beta}$ and
$G_{\alpha\beta\gamma}$ which trivialize the tensorial
supergravity down to either flat tensorial superspace or the
supergroup manifold $OSp(1|n)$.

In the case of the $N=1$, $D=3$ supergravity (where $n=2$) $SL(2)$
is isomorphic to $Sp(2)$, the constraints
(\ref{Tsl1})--(\ref{Tsl2}) are off the mass shell and the
trivialization does not occur. The supergravity equations of
motion are obtained by putting
\begin{equation}\label{RG=0}
R_{\alpha\beta}=0, \qquad G_{\alpha\beta\gamma}=0\,,
\end{equation}
or in the case of AdS
\begin{equation}\label{RG=C}
R_{\alpha\beta}=-{\varsigma\over 2}C_{\alpha\beta}, \qquad
G_{\alpha\beta\gamma}=0\,.
\end{equation}
These equations imply that pure $N=1$, $D=3$ supergravity is
non--dynamical, since its torsion and curvature vanish \cite{D=3}.

Also in the case of $n>2$ the equations (\ref{RG=0}) or
(\ref{RG=C}) single out, respectively, the flat or $OSp(1|n)$
vacuum solution of the tensorial supergravity constraints.

\def\theequation{\arabic{section}.\arabic{equation}}
\setcounter{equation}0

\section{The scalar superfield equation in a tensorial supergravity background}

In the previous sections we have derived the constraints of
tensorial supergravity from the requirement of the
$\kappa$--symmetry of the ``preonic'' superparticle. The
supergravity constraints can also be obtained (see
\cite{SG80,1001} for the ordinary case) by requiring that in
curved superspace there exist (super)field representations of
(generalized) supersymmetry similar to those in flat superspace.
In our case of flat tensorial superspace and of $OSp(1|n)$ the
only known representation is described by the scalar superfield
obeying the dynamical equations (\ref{hsSEq}) and (\ref{DD=AdS}),
respectively. So it is natural to consider a curved superspace
generalization of these equations and to analyze which
restrictions on superspace geometry are imposed by its
integrability. Instead of starting again from the most general
structure of tensorial supergeometry, in this section we shall
consider a possible generalization of eqs. (\ref{hsSEq}) and
(\ref{DD=AdS}) in a curved superspace already subject to the
supergravity constraints (\ref{Tb=AdS}), (\ref{Tf=AdS}) and
(\ref{R=AdS}). Interestingly enough, the integrability of the
scalar superfield equation will require the curved superspace
holonomy to be $SL(n)$ and not $GL(n)$.

 A natural generalization of the free superfield equations
(\ref{hsSEq}) and (\ref{DD=AdS}) is
\begin{equation}\label{DDR}
{\cal D}_{[\alpha}{\cal D}_{\beta]}\,\Phi={i\over
2}\,R_{\alpha\beta}\,\Phi.
\end{equation}
One gets eqs. (\ref{hsSEq}) and (\ref{DD=AdS}) from (\ref{DDR}) by
putting $U_{\alpha\beta\gamma}=0$ and $R_{\alpha\beta}=0$ or
$R_{\alpha\beta}=-{\varsigma\over 2}C_{\alpha\beta}$, respectively
\footnote{Note that eq. (\ref{DDR}) resembles a conformally
invariant scalar field equation in a $D=4$ gravitational
background $g^{mn}{\cal D}_m\partial_n\,b(x)={1\over 6} \,
R(x)\,b(x) $, where $R(x)$ is the curvature scalar.}. A more
general form of the scalar superfield equation is discussed in
Appendix B.

Let us now study the integrability of the equations (\ref{DDR}) in
the case of supergravity with the holonomy group $GL(n)$ subject
to the constraints (\ref{Tb=AdS})--(\ref{R=AdS}). To this end we
need the following covariant derivative commutation relations
(where $W_\delta$ is an arbitrary superfield)
\begin{eqnarray}\label{DD=Db}
\{ {\cal D}_\alpha \, , \, {\cal D}_\beta \} &=& 2i
 {\cal D}_{\alpha \beta } \; ,
\\
\label{DbD=D+} [ {\cal D}_{\alpha\beta} \, , \, {\cal D}_\gamma ]
W_{\delta} &=& - 2 R_{\gamma (\alpha } {\cal D}_{\beta )} W_\delta
- i U_{\delta {\alpha\beta}} W_{\gamma} +  F_{\gamma {\delta
(\alpha}} W_{\beta )} +
 {\cal D}_{\gamma} R_{\delta (\alpha}
 W_{\beta )} \;,
\end{eqnarray}
and
\begin{eqnarray}\label{DbDb=D+}
   [ {\cal D}_{\alpha\beta} \, , \, {\cal D}_{\gamma\delta} ] W_{\epsilon}
  &=&  - 4 R_{(\gamma| (\alpha} {\cal D}_{\beta ) |\delta )}  W_{\epsilon}
  + U_{(\alpha |\,  \gamma\delta } {\cal D}_{|\beta)} W_{\epsilon}
  - U_{( \gamma | \,  \alpha\beta}  {\cal D}_{\delta )}  W_{\epsilon} +
  \qquad \nonumber \\
  &+& {\cal D}_{\gamma\delta} R_{\epsilon (\alpha } \, W_{\beta)} -
  {\cal D}_{\alpha\beta}   R_{\epsilon  (\gamma} \, W_{\delta )}
  + {1\over 2}  {\cal D}_{\epsilon} U_{(\alpha |\,  \gamma\delta } \;
  W_{|\beta)} \qquad \nonumber \\   &-&  {1\over 2}  {\cal
  D}_{\epsilon} U_{( \gamma | \,  \alpha\beta} \, W_{\delta )} -
  {1\over 2}  W_{(\beta}{\cal D}_{\alpha )} U_{\epsilon \,  \gamma\delta } \;
   + {1\over 2}  W_{(\delta } {\cal D}_{ \gamma )} U_{\epsilon \,
   \alpha\beta} \, \,,
   \end{eqnarray}
where it is implied that the indices $(\alpha\beta)$ as well as
$(\gamma\delta)$ are symmetrized  with the unit weight
($A_{\a\b}=A_{(\a\b)}+A_{[\a\b]}$).
 Acting on (\ref{DDR}) with ${\cal D_\gamma}$ and using
(\ref{DD=Db}) and (\ref{DbD=D+}) we get the non--linear
counterpart of the fermionic equations (\ref{hsEqf0}) and
(\ref{fsp})
\begin{eqnarray}\label{fsg}
{\cal D}_{\a[\b}{\cal D}_{\g]}\Phi={1\over 2}\left(R_{\a[\g}{\cal
D}_{\b]}\Phi+R_{\b\g}{\cal D}_\a\Phi \right) -{\Phi\over 6}\,{\cal
D}_{[\beta} R_{\gamma]\alpha}+{\Phi\over 6}\,{\cal D}_{\alpha}
R_{\beta\gamma}\,.
\end{eqnarray}

Acting on (\ref{fsg}) with ${\cal D}_\delta$ and using the
commutation relations (\ref{DD=Db}) and  (\ref{DbD=D+}) we get the
non--linear counterpart of the bosonic equations (\ref{hsEqb0})
and (\ref{bsp})
\begin{eqnarray}\label{bsg}
{\cal D}_{\a[\b}{\cal D}_{\g]\d}\Phi&=&{1\over
2}\left(R_{\b\g}{\cal D}_{\a\d}-R_{\a[\b}{\cal
D}_{\g]\d}-R_{\d[\b}{\cal D}_{\g]\a} \right)\Phi +{1\over
4}\left(R_{\a\d}R_{\b\g}-R_{\a[\b}R_{\g]\d}\right)\Phi\nonumber\\
& +&{i\over 6}{\cal D}_\alpha R_{\beta\gamma}{\cal D}_\delta
\Phi-{i\over 6}{\cal D}_{[\beta} R_{\gamma]\alpha}{\cal D}_\delta
\Phi +{i\over 2}{\cal D}_{\alpha} R_{\delta[\beta}{\cal
D}_{\gamma]} \Phi-{i\over 2}{\cal D}_{[\beta}\Phi\,{\cal
D}_{\gamma]}
 R_{\alpha\delta}\nonumber\\
&+&{\Phi\over 6} (3{\cal D}_{\alpha[\beta}\,R_{\gamma]\delta}
-{{i}} {\cal D}_\delta {\cal D}_\alpha R_{\beta\gamma}+ {{i}}
{\cal D}_\delta {\cal
D}_{[\beta} R_{\gamma]\alpha})\\
&+&U_{[\beta\gamma]\alpha} {\cal D}_\delta\Phi+{1\over
2}U_{(\gamma\alpha)\delta} {\cal D}_\beta\Phi-{1\over
2}U_{(\beta\alpha)\delta }{\cal D}_\gamma\Phi+{1\over
2}U_{\delta\alpha[\beta} {\cal D}_{\gamma]}\Phi\,.\nonumber
\end{eqnarray}
On the other hand, as a consequence of the constraints
(\ref{Tb=AdS}) and (\ref{Tf=AdS}) the `antisymmetrized' commutator
of the bosonic covariant derivatives (\ref{DbDb=D+}) acting on a
scalar superfield has the following form
\begin{eqnarray}\label{compare1}
{1\over 2} [{\cal D}_{\alpha [\beta}, {\cal D}_{\gamma ]\delta}]
\, \Phi={1\over 2}({\cal D}_{\alpha[\beta} {\cal D}_{\gamma
]\delta}+{\cal
D}_{\delta[\beta} {\cal D}_{\gamma ]\alpha})\Phi & \nonumber\\
= -{1\over 2} (R_{\alpha [\beta} {\cal D}_{\gamma ]\delta }\Phi +
R_{\delta [\beta} {\cal D}_{\gamma ]\alpha }\,\Phi-
R_{\beta\gamma} {\cal D}_{\alpha \delta}\,\Phi)&+ {1\over 2} \left
(U_{[\beta\gamma](\alpha} {\cal D}_{\delta)} \Phi -
U_{(\alpha\delta)[\beta}{\cal D}_{\gamma]}\Phi\right)\,.
\end{eqnarray}
 From (\ref{compare1}) it follows that for the equation (\ref{DDR})
to be consistent, the right hand side of the equation (\ref{bsg})
symmetrized in $\alpha$ and $\delta$ must coincide with the right
hand side of (\ref{compare1}). This results in the equation
\begin{eqnarray}\label{compare2}
0 &=&  U_{[\beta\gamma](\alpha}\, {\cal D}_{\delta)} \Phi -{i\over
3}\, {\cal D}_{(\alpha} \Phi {\cal D}_{\delta) }R_{\beta\gamma } -
{i\over 3} {\cal
D}_{[\beta}R_{\gamma ](\alpha}  \, {\cal D}_{\delta)} \Phi  \nonumber \\
&+&  U_{ (\alpha \; \delta) [\beta }\, {\cal D}_{\gamma ]} \Phi +
\, {\cal D}_{[\gamma} \Phi \,U_{\beta ] \, \alpha \delta} + {\cal
D}_{(\alpha}R_{\delta )[\beta }
 \, {\cal D}_{\gamma ]} \Phi \nonumber \\
&+&  {{\Phi}\over {12}} \left( {\cal D}_{\alpha \delta }
R_{\beta\gamma} + 3  {\cal D}_{\alpha [\beta } R_{\gamma]\delta} +
i {\cal D}_{\alpha }  {\cal D}_{[\beta } R_{\gamma]\delta}+\alpha
\leftrightarrow \delta
  \right) \; .
\end{eqnarray}
The above equation is identically satisfied in the case of the
tensorial supergravity with the holonomy group $SL(n)$ described
in Subsection 3.4. Indeed, the first and the second line in
(\ref{compare2}) vanish in virtue of  eq. (\ref{U=W+DR}), while
the last line coincides with the left hand side of eq.
(\ref{compare3}).

Thus, the scalar superfield can consistently propagate in any
tensorial supergravity background with $SL(n)$ holonomy.
Peculiarities of the coupling of a scalar superfield to $N=1$,
$D=3$ supergravity are briefly discussed in Appendix C.

\def\theequation{\arabic{section}.\arabic{equation}}
\setcounter{equation}0
\section{Generalized Weyl invariance of the tensorial supergravity
constraints and conformally related supermanifolds}

Let us now proceed with studying the properties of the tensorial
supergravity constraints and looking for their general solution in
terms of an unconstrained superfield. To this end consider the
following transformations of the supervielbeins and
superconnection of a tensorial superspace
\begin{eqnarray}\label{Wa}
E^{\prime\alpha\beta}&=&E^{\alpha\beta}\,,\nonumber\\
E^{\prime\alpha}\; &=&E^{\alpha}+E^{\alpha\beta}W_\beta\\
\Omega_{\beta}^{\,\prime\,\alpha}&=&
\Omega_{\beta}{}^{\alpha}-iE^{\alpha}W_\beta
-E^{\alpha\gamma}({\cal D}_\gamma\,W_\beta+iW_\gamma W_\beta)\,,
\nonumber
\end{eqnarray}
where $W_\alpha$ is an arbitrary spinor superfield. Then, as one
can check, the form of the supergravity torsion and curvature
(\ref{Tb=AdS})--(\ref{R=AdS}) remain intact when the transformed
$R^\prime_{\alpha\beta}$ and $U^\prime_{\alpha\beta\gamma}$ are
defined as
\begin{eqnarray}\label{RU}
R^\prime_{\alpha\beta}&=&R_{\alpha\beta}-{\cal
D}_{[\alpha}\,W_{\beta]}- {i\over 2}\,W_\alpha W_\beta\,,\nonumber\\
U^\prime_{\alpha\beta\gamma}&=&U_{\alpha\beta\gamma}+{\cal
D}_{\beta\gamma}\,W_{\alpha}-W_{(\gamma} \,{\cal
D}_{\beta)}\,W_{\alpha}\,.
\end{eqnarray}

As a result, the main superfields $R^\prime_{\alpha\beta}$ and
$U^\prime_{\alpha\beta\gamma}$ satisfy the constraints
(\ref{DU=DR})--(\ref{5/2}) provided that $R_{\alpha\beta}$ and
$U_{\alpha\beta\gamma}$ solve them and vice versa.

Thus the solutions of the supergravity constraints
(\ref{DU=DR})--(\ref{5/2}) form classes of equivalence whose
members are related by the transformations (\ref{Wa})--(\ref{RU}).
These can be regarded as a kind of generalized super--Weyl
transformations which reduce to proper Weyl transformations when
$W_\alpha=-i{\cal D}_\alpha\,W(Z)$ with $W(Z)$ being a scalar
superfield (see e.g. \cite{hp,conflat}).

In particular, when
$R^\prime_{\alpha\beta}=0=U^\prime_{\alpha\beta\gamma}$ correspond
to the flat superspace, eqs. (\ref{RU}) describe a class of
conformally flat tensorial superspaces whose holonomy group is
$SL(n)$ if $W_\alpha=-i{\cal D}_\alpha\,W(Z)$
\begin{eqnarray}\label{RU0}
R_{\alpha\beta}&=&{\cal
D}_{[\alpha}\,W_{\beta]} + {i\over 2}\,W_\alpha W_\beta\,,\nonumber\\
U_{\alpha\beta\gamma}&=&-{\cal
D}_{\beta\gamma}\,W_{\alpha}+W_{(\gamma} \,{\cal
D}_{\beta)}\,W_{\alpha}\,.
\end{eqnarray}

To see this let us calculate the trace of the curvature
(\ref{R=AdS}) with $R_{\alpha\beta}$ and $U_{\alpha\beta\gamma}$
given by eqs. (\ref{RU0}). The trace takes the form
\begin{eqnarray}\label{Rabc=W}
{\cal R}_{\alpha  {\beta\gamma},\, \delta}{}^\delta = - i {\cal
D}_{\beta\gamma} W_\alpha + {\cal D}_{\alpha} {\cal D}_{(\beta}
W_{\gamma )} +2i W_{(\beta}R_{\gamma)\alpha} \; .
\end{eqnarray}
For a generic $W_\alpha$ the trace of the curvature is non--zero,
however it identically vanishes when $W_\alpha=-i{\cal D}_\alpha
W$. Indeed in this case, in virtue of the commutation relations
\begin{eqnarray}\label{DD=Db0}
\{ {\cal D}_\alpha \, , \, {\cal D}_\beta \} &=& 2i
 {\cal D}_{\alpha \beta } \; ,
\\
\label{DbD=D0} [ {\cal D}_{\alpha\beta} \, , \, {\cal D}_\gamma ]
W &=& - 2 R_{\gamma (\alpha } {\cal D}_{\beta )} W \;,
\end{eqnarray}
 we get
\begin{eqnarray}\label{Rabc=DPhi}
{\cal R}_{\alpha  {\beta\gamma},\, \delta}{}^\delta =  -[ {\cal
D}_{\beta\gamma}, {\cal D}_{\alpha}] \, W-2
\,R_{\alpha(\beta}{\cal D} _{\gamma)}\,W \equiv 0 \; .
\end{eqnarray}

A simpler way to arrive at the same conclusion is to calculate the
trace of the connection in (\ref{Wa}), $\Omega_{\alpha}^{\prime \,
\alpha} = \Omega_{\alpha} {}^{\alpha} - i E^\alpha W_\alpha -
E^{\alpha\beta} {\cal D}_{\alpha} W_{\beta}$. With $W_\alpha
=-i{\cal D}_{\alpha}W$ and $\Omega_{\alpha}^{\prime \, \alpha} =0$
this gives $\Omega_{\alpha}{}^{\alpha} = dW$ which implies ${\cal
R}_\alpha{}^\alpha=0$.

In conventional superfield theories, a detailed analysis of which
superspaces among supermanifolds containing $AdS_d\times S^m$  are
superconformally flat has been carried out in \cite{conflat}. For
instance, it was demonstrated that the $N=1$ supersymmetric
$AdS_3$ isomorphic to $OSp(1|2)$ is superconformally flat. This is
also a particular case of the tensorial superspace under
consideration when $n=2$, $R_{\a\b}=-{\varsigma\over
2}\,\epsilon_{\alpha\beta}$ and $U_{\a\b\g}=0$.

In \cite{Dima,Dima03} it has been found that $OSp(1|n)$ are so
called $GL(n)$ flat supermanifolds, i.e. their bosonic
supervielbeins $E^{\a\b}$ are obtained from the flat ones  by a
transformation with a certain $GL(n)$ matrix, while the fermionic
supervielbeins $E^\a$ have a more sophisticated form than that of
(\ref{Wa}). For $n=2$ the two properties, superconformal flatness
and GL--flatness, are equivalent since $GL(2)\sim SL(2)\times R$
and $SL(2)\sim Sp(2)$ is the holonomy group of $OSp(1|2)$. As we
have already discussed a supergroup manifold $OSp(1|n)$  with
$n>2$ has the holonomy group $Sp(n)$ which is smaller than
$SL(n)$, therefore in the generic case the properties of
superconformal flatness and of GL--flatness (which, in the way it
works, preserves $Sp(n)$--holonomy) are not equivalent and hence
do not imply each other.

Indeed, the supergroup manifold $OSp(1|n)$ with $n>2$ is not
superconformally flat. To show this let us recall again that the
main superfields $R_{\a\b}$ and $G_{\a\b\g}$ of $OSp(1|n)$ satisfy
the equations (\ref{RG=C}) and $U_{\a\b\g}=0$ which imply that
\begin{equation}\label{covco}
{\cal D}_{(\a}\,C_{\beta)\gamma}=D_{(\a}\,C_{\beta)\gamma}
-\Omega_{(\a \,\beta)}{}^{\delta}\,C_{\gamma\delta} +
C_{\delta(\a}\,\Omega_{\b)\, \gamma}{}^{\delta} =0\, \quad
\Rightarrow \quad \Omega_{(\a\,
[\beta)}{}^{\delta}\,C_{\gamma]\delta}=0\,.
\end{equation}
Substituting into (\ref{covco}) the superconformally flat form of
the connection (see eq. (\ref{omega}) of Section 5) we arrive at
the condition
\begin{equation}\label{n=2}
(n-2)(n+1)\,D_\a\,W=0,
\end{equation}
from which it follows that the Weyl scalar superfield $W(Z)$ does
not reduce to the constant only when $n=2$. When $n>2$, $W=const$
and thus (\ref{RG=C}) are consistent with (\ref{RU0}) if only
$\varsigma=0$.  Hence, $OSp(1|n)$ with $n>2$ is not
superconformally flat.

As so, in the case, when $R^\prime_{\alpha\beta}=-{\varsigma\over
2}C_{\alpha\beta}$ and $U^\prime_{\alpha\beta\gamma}=0$ which are
that of the supermanifold $OSp(1|n)$, the generalized Weyl
transformations produce a class of tensorial superspaces which are
not superconformally flat but are conformally related to
$OSp(1|n)$ with $R_{\alpha\beta}$ and $U_{\alpha\beta\gamma}$
having the following form
\begin{equation}\label{Rosp}
R_{\alpha\beta}=-{\varsigma\over 2}C_{\alpha\beta}+{\cal
D}_{[\alpha}\,W_{\beta]}+ {i\over 2}\,W_\alpha W_\beta\,,
\end{equation}
\begin{equation}\label{Uosp}
U_{\alpha\beta\gamma}=-{\cal
D}_{\beta\gamma}\,W_{\alpha}+W_{(\gamma} \,{\cal
D}_{\beta)}\,W_{\alpha}\,.
\end{equation}
We should note that though the supermanifold $OSp(1|n)$ we started
with has the holonomy group $Sp(n)$ with respect to which
$C_{\alpha\beta}$ is covariantly constant, the resulted
supermanifolds described by (\ref{Rosp}) and (\ref{Uosp}) have (in
general) the holonomy group $GL(n)$ which reduces to $SL(n)$ when
$W_\alpha=-i{\cal D}_\alpha\,W(Z)$. With respect to the $GL(n)$--
or $SL(n)$--covariant derivative $C_{\alpha\beta}$ is not
covariantly constant. Hence, the equation (\ref{Rosp}) should be
considered as valid in some gauge which reduces $GL(n)$ or $SL(n)$
down to $Sp(n)$. The $GL(n)$ covariant expression for
$R_{\alpha\beta}$ is
\begin{equation}\label{Rospco}
R_{\alpha\beta}=-{\varsigma\over 2}{\cal X}_{\alpha\beta}(Z)+{\cal
D}_{[\alpha}\,W_{\beta]}+{i\over 2}\,W_\alpha W_\beta\,,
\end{equation}
where ${\cal X}_{\alpha\beta}(Z)$ is now an antisymmetric tensor
superfield with $\det {\cal X}_{\alpha\beta}\not =0$. Using a
local $GL(n)$ transformation ${\cal
X}^\prime_{\alpha\beta}(Z)=G_\alpha^{~\alpha'}(Z)\,G_\beta^{~\beta'}(Z)\,{\cal
X}_{\alpha'\beta'}(Z)$  it is always possible to put ${\cal
X}^\prime_{\alpha\beta}(Z)=C_{\alpha\beta}$ and to reduce
(\ref{Rospco}) to (\ref{Rosp}).

\def\theequation{\arabic{section}.\arabic{equation}}
\setcounter{equation}0
\section{Decoupling of higher spin field dynamics from superconformal geometry}
Since the higher spin fields (at least at the linearized level)
are described by the single scalar superfield it is natural to
assume that in order to switch on non--trivial higher spin
interactions the geometry of tensorial supergravity coupled to the
scalar superfield is itself expressed in terms of this single
scalar superfield (see \cite{Misha0304b} for a somewhat similar
assumption in the framework of the unfolded higher spin
formulation). If we restrict ourselves to the class of conformally
flat manifolds or to the class of manifolds conformally related to
$OSp(1|n)$ discussed in the previous Section and assume that the
geometry is expressed in term of a single physical scalar
superfield $\Phi$ we find that the geometry reduces to flat
superspace (or to $OSp(1|n)$ superspace) and the superfield $\Phi$
(in a sense) decouples from the geometry. The reason is in the
generalized super--Weyl invariance of both the supergravity
constraints and the scalar superfield equation (\ref{DDR}).

Let us first discuss the conformally flat case and then the
$OSp(1|n)$ related one.

Using the generic expressions of Section 4 the supervielbeins and
the $SL(n)$ connection of a conformally flat superspace  can be
presented in the following conventional form ({\it cf.}  e.g.
\cite{hp,conflat})
\begin{equation}\label{e}
E^{\a\b}=e^{{2W(Z)}\over n}\,E^{\a'\b'}_0\,L_{\a'}^{~\a}(Z)\;
L_{\b'}^{~\b}(Z)\; , \quad E^\a=e^{{W(Z)}\over
n}\,(E^{\a'}_0-iE^{\a'\b'}_0\,D_{\b'}\,W) \,L_{\a'}^{~\a}(Z),
\end{equation}
\begin{eqnarray}\label{omega}
\Omega^{~\a}_\b=\Omega_{0\,\b}{}^\a+{1\over n}
dW\,\delta_\b{}^\a -L^{-1}{}_{\b}{}^{\b'}\,\left[E^{\a'}_0\,D_{\beta'}
W+E^{\a'\g}_0 (D_{\g\b'}W+{i\over 2}D_\gamma W\,D_{\beta'}
W)\right]L_{\a'}^{~\a}, \quad \\ \Omega^{~\a}_\a \equiv 0\; ,  \qquad
\nonumber
\end{eqnarray}
where $L_{\b}^{~\a}(Z)$ is a matrix of local $SL(n)$
transformations which together with $e^{{W(Z)}\over n}$ form a
$GL(n)$ matrix $G_{\b}^{~\a}=e^{{W(Z)}\over n}L_{\b}^{~\a}$. The
supervielbeins $E^{\a\b}_0$, $E^\a_0$ and the connection
$\Omega^{~~\a}_{0\b}$
 satisfy the constraints of a flat superspace
\begin{equation}\label{flat}
T^{\a\b}_0=-iE^\a_0\wedge E^\b_0, \qquad T^\a_0=0={\cal
R}_0{}^{~\a}_\b
\end{equation}
and $D_{\a\b}$ and $D_\a$ are corresponding covariant derivatives.

In particular, in the `flat' basis
\begin{equation}\label{or}
E^{\a\b}_0=dX^{\a\b}-id\theta^{(\a}\,\theta^{\b)}, \quad
E^\a_0=d\theta^\a\,, \quad \Omega^{~~\a}_{0\b}=0\,,\quad
L_{\b}^{~\a}=\delta_{\b}^{~\a}\,,
\end{equation}
\begin{equation}\label{Dor}
D_{\a\b}=\partial /
\partial X^{\a\b}\equiv \partial_{\a\b}, \quad
D_\a=\partial / \partial \theta^\a+i\theta^\b\,\partial_{\b\a}\,
\end{equation}
and
\begin{equation}\label{calD}
{\cal D}_{\a\b}=e^{-{{2W}\over n}}(D_{\a\b}-iD_{(\a}W\,D_{\b)})+
\Omega_{\a\b} - i e^{-{{W}\over n}}D_{(\a}W\,\Omega_{\b)} \; ,
\quad {\cal D}_\a=e^{-{{W}\over n}}{D}_\a+\Omega_\a\,,
\end{equation}

Using the constraint relations (\ref{Tb=M})--(\ref{R=M}) and
(\ref{Raa=0}) one finds that in the `flat' basis the main
superfields $R_{\alpha\beta}$ and  $U_{\beta{\gamma\delta}}$ have
the form
\begin{equation}\label{DDRW}
R_{\alpha\beta}= \, i\,e^{-{{2W}\over n}}\,\left[{ D}_{[\alpha}{
D}_{\beta]}\,W+{1\over 2}\,{ D}_\alpha W\, { D}_\beta W\right]\,,
\end{equation}
\begin{equation}\label{UWf}
U_{\beta\,
 {\gamma\delta}}=e^{-{{3W}\over n}}
\left[- i {\partial}_{\gamma\delta} { D}_{\beta}
 W + D_{(\g}W\,D_{\delta)}D_{\b}W\right]\,,
\end{equation}
or in the basis of the `curved' covariant derivatives (\ref{calD})
\begin{eqnarray}\label{DDRg}
R_{\alpha\beta}&=& i{\cal D}_{[\alpha}{\cal D}_{\beta]}\,W -
{i\over 2} \,{\cal D}_\alpha W\, {\cal D}_\beta W \nonumber \\ &=&
-{\cal D}_{[\alpha}W_{\beta]} + {i\over 2} \,W_\alpha \,
W_\beta\,,
\end{eqnarray}
 \begin{eqnarray}
 \label{U=DW+WDW}
U_{\beta \, {\gamma\delta}} &=& i {\cal D}_{\gamma\delta} {\cal D}_{\beta}\,W -
{\cal D}_{(\gamma }\,W\,  {\cal D}_{\delta )}{\cal D}_{\beta}\,W \nonumber \\
&=& - {\cal
D}_{\gamma\delta} W_{\beta} + W_{(\gamma }  {\cal D}_{\delta )}
W_{\beta}\; .
\end{eqnarray}
where we have introduced $W_\alpha\equiv -i{\cal D}_\alpha W$ to
compare these expressions with those of Section 4.

In view of a generic reasoning behind the constraint conserving
transformations of the supervielbeins and superconnection given in
Section 4 one can directly check that the main superfields
(\ref{DDRW})--(\ref{U=DW+WDW}) identically satisfy the constraints
(\ref{DU=DR})--(\ref{Raa=0}) and (\ref{5/2})  of tensorial
supergravity with $SL(n)$ holonomy, which can be checked directly.

Now our assumption that the geometry depends only on the scalar
superfield $\Phi$ implies that $W$ becomes a scalar function of
$\Phi$, $W=W(\Phi)$, and using this (physical) scalar superfield
$W(\Phi)$, we are allowed to perform the Weyl transformation
(\ref{Wa}) and get for the transformed superfields $R'_{\alpha
\beta}= 0 = U'_{\alpha
\beta \gamma}$, i.e. flat superspace \footnote{Note that if we {\sl
formally}
put to zero only $R_{\a\b}$ while keeping $U_{\alpha
\beta \gamma}$ in the form (\ref{U=DW+WDW}) we get
$$
R_{\alpha\beta}= 0 \qquad \Rightarrow \qquad { D}_{[\alpha}{
D}_{\beta]}\,W+{1\over 2}\,{ D}_\alpha W\, { D}_\beta W =0 \; .
$$
This equation reduces to the free scalar superfield equation
(\ref{hsSEq}) upon the field redefinition $W=2\mathrm{ln} \Phi$,
or better $W=2\mathrm{ln} (\Phi + a)$ with an arbitrary constant
$a>0$. So one might think that at least free higher spin dynamics
is intrinsically encoded in superconformally flat tensorial
geometry, but this is not the case since using the super--Weyl
transformations with a parameter satisfying the free scalar
superfield equation one can put $U_{\alpha
\beta \gamma}=0$ and arrive in flat superspace with no dynamics.}. Thus
the superfield $\Phi$ decouples from supergravity, and the most
general form of the scalar superfield equation which one may
construct in such a case is
\begin{equation}\label{dynamic}
D_{[\a} D_{\b]}\,\Phi={\cal X}_{\a\b}(\Phi)\,,
\end{equation}
where ${\cal X}_{\a\b}(\Phi)$ is an antisymmetric tensor which
depends on $\Phi$ and its derivatives. ${\cal X}_{\a\b}(\Phi)$
must satisfy an integrability condition (see (\ref{selfcX}) of the
Appendix B, where one should put ${\cal D}_\alpha=D_\alpha$,
${\cal D}_{\alpha\beta}=\partial_{\alpha\beta}$  and
$R_{\alpha\beta}=0$).

If, for example, we choose ${\cal X}_{\a\b}(\Phi)=-f'(\Phi)D_\a
\Phi D_\b \Phi$, where $f(\Phi)$ is an arbitrary function and
$f'(\Phi)={{df}\over {d\Phi}}$, the eq. (\ref{dynamic}) takes the
form
$$
D_{[\a} D_{\b]}\,\Phi+f'(\Phi)D_\a \Phi D_\b \Phi=0\,,
$$
which upon the field redefinition $\tilde\Phi=const\cdot\int
d\Phi\, e^{f(\Phi)}$ (i.e. ${{d\tilde\Phi}\over{d\Phi}}=const\cdot
e^{f(\Phi)}$) reduces to the free scalar superfield equation
(\ref{hsSEq}).

If there exists a more general ${\cal X}_{\a\b}(\Phi)$ satisfying
the consistency condition (see eq. (\ref{selfcX}) of Appendix B),
the equation (\ref{dynamic}) would describe a non--linear dynamics
of a self--interacting scalar superfield $\Phi(Z)$. Since, as we
have explained in the Introduction, $\Phi(Z)$ contains only the
linearized field strengths of the higher spin fields and not their
potentials, such a non--linear dynamics of higher spin fields
would not include minimal coupling terms which require potentials
or connections. It would contain only terms constructed of higher
orders of the higher spin field strengths. As a result, the
non--linear model obtained in this way would be analogous in a
certain sense to the abelian Dirac--Born--Infeld theory.

Let us now discuss the case of the tensorial manifolds conformally
related to $OSp(1|n)$. The consideration follows the same lines as
in the case of the conformally flat manifolds and the result is
that in terms of the $OSp(1|n)$ covariant derivatives the main
tensor fields describing their geometry have the following form
\begin{equation}\label{DDRWo}
R_{\alpha\beta}=\, i\,e^{-{{2W}\over n}}\,\left[\, i{\varsigma\over
2}C_{\alpha\beta}+{ \nabla}_{[\alpha}{ \nabla}_{\beta]}\,W+{1\over
2}\,{ \nabla}_\alpha W\, { \nabla}_\beta W\right]\,,
\end{equation}
\begin{equation}\label{UWfo}
U_{\beta
 {\gamma\delta}}=e^{-{{3W}\over n}}
\left[- i {\nabla}_{\gamma\delta} { \nabla}_{\beta}
 W+
\nabla_{(\g}W\,\nabla_{\delta)}\nabla_{\b}W\right]\,.
\end{equation}
With such a definition of $R_{\alpha\beta}$ and $U_{\beta
 {\gamma\delta}}$ the spin connection of this tensorial superspace
 is $SL(n)$--valued.

 Again one can now perform  the field redefinitions (4.1)
to end with the $OSp(1|n)$ geometry. As in the case of the
conformally flat superspace the field $W$ disappears from the
transformed torsion and curvature and eventually we can impose on
$\Phi$  the linear scalar superfield equation \footnote{Note that,
as in the superconformally flat case this equation can be obtained
from eq. (\ref{DDRWo}) by a formal trick, namely by putting in
(\ref{DDRWo}) $R_{\a\b}= - {\varsigma\over 2}C_{\alpha\beta}\,
exp\{(1+4/n)\, W/2 \}$ and making in the resulting equation
$$ \nabla_{[\a} \nabla_{\b]} W +
{1\over 2} \nabla_{\a}W\, \nabla_{\b} W = - {i\varsigma\over
2}C_{\alpha\beta} \left(1- e^{-{W\over 2}}\right)\;
$$
the field redefinition $W= 2\, \mathrm{ln} \left({\Phi + a\over
a}\right) \; ,~  a>0\; .$ }
$$
\left(\nabla_{[\alpha} \nabla_{\beta]} + i{\varsigma\over 4}
C_{\alpha\beta} \right) \Phi(X,\theta)=0 \; .
$$

As in the case of the flat tensorial superspace a problem for
future study is to understand if the integrability conditions
discussed in Appendix B allow for the existence of more general,
non--linear scalar superfield equation on $OSp(1|n)$ in the form
\begin{equation}\label{dynamico}
\nabla_{[\a} \nabla_{\b]}\,\Phi={\cal X}_{\a\b}(\Phi)\,.
\end{equation}

{} From the above discussion we can conclude that to construct
non--linear higher spin equations involving in a non--trivial way
higher spin field potentials  one should have at his disposal more
general tensorial superspaces than superconformally flat manifolds
or manifolds conformally related to $OSp(1|n)$. However, as we
shall demonstrate in the next section, the superconformal
tensorial superspaces are the general solution of the supergravity
constraints (at least locally, or when the first cohomology of the
tensorial superspace is trivial).

\def\theequation{\arabic{section}.\arabic{equation}}
\setcounter{equation}0
\section{The general solution of the tensorial
supergravity constraints}

 Let us show that (up to topological
subtleties) the superconformally flat and $OSp$ related geometries
studied in Sections 5 and 6 form the general solution of the
tensorial supergravity constraints (\ref{Tb=AdS}), (\ref{Tf=AdS})
and (\ref{R=AdS}). To this end consider a weak deviation of
tensorial supergeometry from the `vacuum' solutions, namely, from
the flat superspace (\ref{RG=0}), and from the  supergroup
manifold $OSp(1|n)$ (\ref{RG=C}).

In the weak superfield approximation over the flat superspace the
main superfields describing such a curved tensorial superspace are
considered to be infinitesimal of order one, $R_{\alpha
\beta}=o(1)$ and $U_{\alpha \; \beta\gamma}= o(1)$. The
constraints (\ref{DU=DR}), (\ref{DU=DF}) and (\ref{5/2}) on these
superfields should be satisfied order by order and in particular
in the linear approximation for the infinitesimal quantities of
order one. In this approximation we ignore the connection terms in
the covariant derivatives (which thus become those of the flat
superspace) and drop the second order terms in eq. (\ref{5/2}).
Then eq. (\ref{5/2}) takes the form
\begin{equation}\label{5/2-0}
\partial_{\a \b} U_{\g\delta \sigma} -
\partial_{\delta \sigma} U_{\gamma \a \b} =0.
\,
\end{equation}
As a consequence of the Poincar\'e lemma its general solution is
\begin{equation}\label{U=dPsi0}
U_{\gamma \a \b} = -\partial_{\a\b} \Psi_\gamma \, .
\end{equation}
Now in the linear approximation eq. (\ref{DU=DR}) reduce to
\begin{eqnarray}\label{DU=DR-0}
\partial_{\gamma\delta} R_{\alpha\beta} =
 D_{[\alpha|}\partial_{\gamma\delta} \Psi_{|\beta]} =
\partial_{\gamma\delta} D_{[\alpha}\Psi_{\beta]}
 \; .
\end{eqnarray}
Its general solution is
\begin{eqnarray}\label{R=DPsi-0}
R_{\alpha\beta} = D_{[\alpha}\Psi_{\beta]} + a_{\alpha\beta}
  \; , \qquad \partial_{\gamma\delta}a_{\alpha\beta}=0\; , \qquad
  a_{\alpha\beta}=- a_{\b\a} = o(1)\; ,
\end{eqnarray}
where $a_{\alpha\beta}$ is independent of $X^{\a\b}$. In the
simplest case when  $a_{\a\b}$ is a constant matrix it can be
absorbed by  $\Psi_{\beta}$ if one performs the following
redefinition $\Psi_{\beta} \rightarrow \Psi_{\beta} +
\theta^\gamma a_{\gamma\beta}$. If $a_{\a\b}$ is a generic
polynomial in $\theta$, the solution (\ref{R=DPsi-0}) breaks the
$GL(n)$ symmetry and supersymmetry of the original system of
supergravity constraints.

 When $a_{\alpha\beta}=0$, eqs. (\ref{5/2-0}) and
(\ref{R=DPsi-0}) describe a weak superfield approximation  of the
superconformally flat geometry (\ref{DDRg}) and (\ref{U=DW+WDW})
with  $W_\a = \Psi_\a = o(1)$. Extending the above analysis to
higher orders in the superfields we find that the superconformally
flat geometry is the general solution of the constraints on
tensorial supergravity which is continuously related to the flat
superspace vacuum.

Let us now consider curved tensorial superspaces with the holonomy
group $GL(n)$ or $SL(n)$ whose geometry weakly differs from the
`vacuum' superspace $OSp(1|n)$ (\ref{RG=C}). In the weak field
approximation the superfield $U_{\alpha \beta\gamma}$ is
infinitesimal of order one
 $U_{\alpha \beta\gamma}= o(1)$, while
$R_{\alpha \beta}=-{\varsigma\over 2}\, C_{\a\b}+r_{\a\b}$ is of
order zero, with $r_{\a\b}$ being infinitesimal of order one .
Note also that the covariant derivatives of $R_{\alpha
\beta}$ are infinitesimal of order one ${\cal D}R_{\alpha
\beta}=o(1)$\footnote{We should stress that one is not obliged to use the
explicit form of the $OSp$--vacuum solution involving the constant
matrix $C_{\a\b}$ which breaks manifest $GL(n)$ or $SL(n)$ gauge
invariance. All the consideration can be carried out in a $GL(n)$
covariant fashion by using two properties of the `near--OSp'
superfields: $R_{\a\b}$ should be a non--degenerate matrix of
order zero and ${\cal D}R_{\a\b}$ is infinitesimal of order one.}.
More explicitly, in the linear approximation
\begin{equation}\label{Dr}
{\cal D}R_{\alpha
\beta}=(\nabla+\Omega)R_{\a\b}=-{\varsigma\over 2}{\cal
D}\,C_{\a\b}+\nabla\,r_{\a\b}=\varsigma\,\Omega_{[\a}
^{~~\gamma}C_{\b]\g} + \nabla\,r_{\a\b}=o(1)\,,
\end{equation}
where $\nabla$ are the covariant derivatives satisfying the
$osp(1|n)$ superalgebra (\ref{sp}) and (\ref{osp1})  (note that
$\nabla\,C_{\a\b}=0$), and $\Omega_{\a}^{~~\b}$ is an order one
deviation of the $GL(n)$ (or $SL(n)$) connection of the curved
superspace from the $Sp(n)$ connection of the supermanifold
$OSp(1|n)$.

In the linear approximation the equation (\ref{5/2}) takes the
form
\begin{equation}\label{5/2osp}
{\nabla}_{\a \b} U_{\g\delta \sigma} - {\nabla}_{\delta \sigma}
U_{\gamma \a \b} -\varsigma \, U_{\g \a(\sigma} C_{\delta)\b} -
\varsigma\, U_{\g\b(\sigma} C_{\delta)\a} = 0\,.
\end{equation}
Its general solution is
\begin{equation}\label{U=dPsi1}
U_{\gamma \a \b} =- {\nabla}_{\a\b} \Psi_\gamma \, .
\end{equation}
Using (\ref{U=dPsi1}) and (\ref{DbD=D+}) in the weak superfield
approximation one finds that eq. (\ref{DU=DR}) reduces to
\begin{equation}\label{D(R+DW)=}
{\cal D}_{\g\d} (R_{\a \b} - {\cal D}_{[\a} \Psi_{\b ]})= - 2
R_{[\a|(\g} \, {\cal D}_{\d )}\Psi_{|\b ]} \, ,
\end{equation}
and in view of (\ref{Dr})
\begin{equation}\label{DC=}
 \varsigma\,
\Omega_{\gamma\delta\, [\a}{}^\e C_{\beta]\e }  - \varsigma\,
C_{[\a|(\g} \, {\nabla}_{\d )}\Psi_{|\b ]} =
 - {\nabla}_{\g\d}  (r_{\alpha\beta} - {\nabla}_{[\a} \Psi_{\b ]}) \, .
\end{equation}
One easily sees that a particular solution of (\ref{DC=}) is
\begin{eqnarray}\label{solosp}
r_{\alpha\beta} &=& {\nabla}_{[\a} \Psi_{\b ]}\quad \rightarrow
\quad  R_{\alpha
\beta}=-{\varsigma\over 2}\, C_{\a\b}+{\nabla}_{[\a} \Psi_{\b
]}\,,\\
\Omega^{~\b}_{\a}&=&(\Omega_{_{OSp}})^{~\b}_{\a}+
iE^{\alpha}_{_{OSp}}\,\Psi_\beta
-E^{\alpha\gamma}_{_{OSp}}{\nabla}_\gamma\,\Psi_\beta \,.\nonumber
\end{eqnarray}
The solutions (\ref{U=dPsi1}) and (\ref{solosp}) are the
linearized version of (\ref{Wa}), (\ref{Rosp}) and (\ref{Uosp})
which describe the tensorial superspaces conformally related to
the supermanifold $OSp(1|n)$.

To understand whether a more general solution of the equation
(\ref{DC=}) exists notice that the main superfield $U_{\a\b\g}$
expressed as in (\ref{5/2osp}) can be put to zero by using
generalized super--Weyl transformations (\ref{Wa}) with
$W_\a=-\Psi_\a$. Equivalently, one can simply put $\Psi_\a=0$ in
eqs. (\ref{5/2osp}), (\ref{D(R+DW)=}) and (\ref{DC=}). We thus
exclude from further consideration the superconformal solution
already found above. Then (\ref{D(R+DW)=}) and (\ref{DC=}) reduce
to
\begin{equation}\label{u=0}
{\cal D}_{\g\d}R_{\a\b}=0 \quad \rightarrow \quad
\varsigma\,\Omega_{\gamma\delta\, [\a}{}^\e C_{\beta]\e }=-
{\nabla}_{\g\d}  r_{\alpha\beta} \; .
\end{equation}
If we restrict the consideration to superspaces of $SL(n)$
holonomy then $R_{\a\b}$ must also satisfy the condition
\begin{equation}\label{DR=0}
{\cal D}_{(\g}\,R_{\b)\a}=0,
\end{equation}
which follows from the fact that for the spaces of $SL(n)$
holonomy $U_{\a\b\g}=G_{\a\b\g}+{{2i}\over 3}{\cal
D}_{(\g}\,R_{\b)\a}$ (see Subsection 3.4) and in the case under
consideration $U_{\a\b\g}=0$.

It can be shown, using the commutation relations (\ref{DbD=D+})
for the covariant derivatives and assuming $R_{\a\b}$ to have the
inverse matrix, that a stronger condition holds ${\cal
D_\g}\,R_{\b\a}=0$.  Then together with (\ref{u=0}) this implies
that $R_{\a\b}$ is covariantly constant ${\cal D}\,R_{\a\b}=0$
\footnote{Indeed, (\ref{DR=0}), (\ref{u=0}) and (\ref{DbD=D+})
with $U_{\b\; \g\d}=0$ imply ${\cal D}R_{\a(\g} \, R_{\d )\b} +
{\cal D}R_{\b(\g} \, R_{\d )\a}=0$ which also can be written in
the form ${\cal D}(R_{\a(\g} \, R_{\d )\b})=0$. In the case with
an invertible $R_{\a\b}$ one multiplies this equation by $R^{-1\,
\e\a} R^{-1\, \k \b}$ to arrive at $(R^{-1}{\cal D}R)_{(\g}^{\;
(\e} \, \d_{\d)}^{\; \k)} =0$ which implies $(R^{-1}{\cal
D}R)_{\g}^{\; \e}=0$ and, hence, ${\cal D}R_{\b\g}=0$.} whose
integrability (in view of the constraints
(\ref{Tb=AdS})-(\ref{R=AdS}) on the torsion and curvature) forces
the tensorial superspace to be the supergroup manifold $OSp(1|n)$.

We have thus shown that the general solution of the tensorial
supergravity constraints are the superspaces conformally related
to flat superspace or supergroup manifold $OSp(1|n)$.

\def\theequation{\arabic{section}.\arabic{equation}}
\setcounter{equation}0

\section{Conclusion and discussion}
The main results of this article are the following
\begin{itemize}
\item
we have found simple free equations of motion of a scalar
superfield propagating in flat tensorial superspace (\ref{hsSEq})
and in the supergroup manifold $OSp(1|n)$ (\ref{DD=AdS}) which in
the case of $n=4$ describe the infinite set of $OSp(1|8)$
invariant free higher spin field equations in flat $D=4$  and
$AdS_4$ space--time, respectively; in the cases of $n=8$ and
$n=16$, which correspond to $D=6$ and $D=10$ space--time, these
equations describe conformally invariant higher spin fields with
self--dual field strengths (work in progress);
\item
the geometry of curved tensorial superspaces has been introduced
and corresponding supergravity constraints have been obtained from
the requirement of the $\kappa$--symmetry of superparticle
dynamics in the tensorial supergravity background; the superfield
structure of the tensorial supergravity has been shown to be a
generalization of $N=1$, $D=3$ supergravity;
\item
 A `no--go' result is that the class of the
superconformally flat and $OSp(1|n)$--related  superspaces is the
general solution of the constraints of tensorial supergravity with
$GL(n)$ or $SL(n)$ holonomy which are required by the
$\kappa$--symmetry of the {\it $GL(n)$--invariant} tensorial
superparticle.

As we have shown, the geometry of these superspaces is trivial in
the sense that it cannot produce `minimal--like' interactions of
higher spin fields.
\end{itemize}

During work on this project we have also analyzed the possibility
of constructing a tensorial super--Yang--Mills theory and its
coupling to the scalar superfield and have not found a nontrivial
model of this kind which possess manifest $GL(n)$ or $SL(n)$
symmetry and non--manifest $OSp(1|2n)$ generalized superconformal
symmetry. One can thus assume that, surprisingly enough, the
scalar superfield is the only dynamical object in the tensorial
superspace of this kind. This is similar to the unfolded higher
spin field dynamics of Vasiliev where at the linearized level all
physical degrees are contained in a scalar field (zero form).
Since interactions of higher spin fields break conformal
invariance one should look for tensorial superfield models in
which the generalized superconformal group $OSp(1|2n)$ and a
corresponding structure group $GL(n)$ or $SL(n)$ are
(spontaneously) broken down to an appropriate subgroup (or
realized non--linearly with an appropriate linearly realized
subgroup). The unbroken/linearly realized subgroup of $GL(n)$ or
$SL(n)$ should presumably be the Lorentz group $SO(1,D-1)$ of the
associated D--dimensional subspace--time of the tensorial
superspace.

 One might hope that in such models the tensorial
supergravity constraints are less restrictive.

Note that in the unfolded formulation of non--linear higher spin
dynamics conformal symmetry is spontaneously broken by doubling
auxiliary (spinor or vector) variables and introducing
Goldstone--like fields which acquire non--zero vacuum expectation
values \cite{V01s}.

Our results suggest that for tensorial supergravity to be a
relevant geometrical framework for the formulation of non--linear
dynamics of higher spin fields in a way which would be somewhat
alternative to the unfolded higher spin dynamics
\cite{unfold,V99,V01} one should enlarge the superspace with
additional coordinates, for example, by keeping in the non--linear
construction the auxiliary commuting spinor variables which were
used to construct the superparticle action in the tensorial
superspace and which entered the `preonic' field equations
(\ref{hsEqbl}) and (\ref{ysp}). In this respect let us conclude
with the following comment. As we have already mentioned, most of
the known approaches to the description of higher spin theories
use additional variables, like vector variables \cite{Misha0304b}
or bosonic spinor variables (see e.g.
\cite{unfold,V99,V01,Misha03s} and refs. therein). The
construction of non--linear higher spin equations based on the
unfolded formulations requires the doubling of the auxiliary
variables {\it of the same kind} \cite{Misha0304b} and
(spontaneous) breaking of conformal symmetry.

When higher spin theories are formulated in a tensorial space or
superspace, as discussed in this paper, in addition to the
ordinary space--time coordinates $x^m$ one introduces auxiliary
tensorial variables ($y^{mn}=-y^{nm}$ for $D=4$).  Higher spin
field equations can be regarded as those which describe the
physical states of a first quantized particle. To construct an
appropriate classical mechanics of this particle one also needs
bosonic spinor variables $\lambda_\alpha$. The quantization of
this particle mechanics \cite{BLS99} produces the field equation
(\ref{hsEqbl}) or (\ref{hsEqby}). Then, in the free field theory
case one can consistently eliminate the dependence of the wave
functions on either the tensorial variables $y^{mn}$ and recover
the unfolded formulation \cite{V01s,Dima,Dima03}, or on the
spinorial variables $\lambda_\alpha$ and get the higher spin field
equations (\ref{hsEqb0}) in tensorial spaces \cite{V01s,V01c}.
Thus, in view of the above remark on `doubling' one can assume
that the formulation of the non--linear dynamics of higher spin
fields in the framework of tensorial SYM or supergravity may
require both the tensorial and spinorial auxiliary variables. In
this perspective the superfield generalization of the `preonic'
equations (\ref{preonsusy}) and (\ref{adspreonsusy}) may play a
special role.

\bigskip

{\bf Acknowledgments}. We would like to thank, Xavier Bekaert,
Nathan Berkovits, Evgeny Ivanov, Mirian Tsulaia and Mikhail
Vasiliev for useful discussions. This work was partially supported
by the research grants BFM2002-03681 from the Ministerio de
Educaci\'on y Ciencia and from EU FEDER funds, by the grant N 383
of the Ukrainian State Fund for Fundamental Research, by the INTAS
Research Project N 2000-254 and by the European Community's Human
Potential Programme under contract HPRN-CT-2000-00131 ``Quantum
Spacetime''.

\def\theequation{A.\arabic{equation}}
\setcounter{equation}0
\section*{Appendix A. Spinor superfield equations}

One may ask whether it is possible instead of considering the
scalar superfield equations (\ref{hsSEq}) to incorporate component
eqs. (\ref{hsEqb0}) and (\ref{hsEqf0}) into equations for a spinor
superfield whose leading component is the fermionic field
$f_\alpha(X)$? The answer to this question is positive, although
one should require the spinor superfield to obey a set of two
equations
\begin{eqnarray}\label{hsSEqS1}
D_{[\alpha}\Psi_{\beta ]} (X, \theta )= 0 \; , \qquad \\
\label{hsSEqS2}
\partial_{\alpha[\beta} \Psi_{\gamma ]} (X, \theta ) =0 \; .
\qquad
\end{eqnarray}
Indeed, in virtue of eq. (\ref{hsSEqS1}), one finds that
$D_{[\beta} D_{\gamma]} \Psi_\alpha = 2i\partial_{\alpha [\beta}
\Psi_{\gamma]}$. Then, because of (\ref{hsSEqS2}),
\begin{eqnarray}\label{hsSEqS3}
D_{[\alpha}D_{\beta ]} \Psi_\alpha (X, \theta ) = 0 \; .
\end{eqnarray}
Eq. (\ref{hsSEqS3}) implies that the spinor superfield
$\Psi_\alpha (X, \theta )$ contains only two non--zero components
\begin{eqnarray}\label{Psi=1+2}
\Psi_\alpha (X, \theta ) = f_\alpha (X) + \theta^\beta k_{\beta
\alpha}(X) \;  .
\end{eqnarray}
Imposing eq. (\ref{hsSEqS1}), one finds that the bosonic
spin-tensor $k_{\beta\alpha}$ is symmetric,
$k_{\beta\alpha}=k_{\alpha\beta}$, and the fermionic field
$f_\alpha (X)$ obeys the equations (\ref{hsEqf0}). The same
equations follow from eq. (\ref{hsSEqS2}), which also implies that
$
\partial_{\alpha [\beta } k_{\gamma ] \delta} = 0$. The latter can be
decomposed into
\begin{eqnarray} \label{hsSEqS4}
\partial_{\alpha  [\beta }
k_{\gamma ] \delta }+\partial_{\delta  [\beta }
k_{\gamma ]\alpha} = 0 \;  , \\
\label{hsSEqS5}
\partial_{\alpha [\beta }
k_{\gamma ] \delta } - \partial_{\delta  [\beta } k_{\gamma ]
\alpha }= 0 \;  .
\end{eqnarray}
Eqs. (\ref{hsSEqS4}) are actually a kind of Bianchi identities
which imply that the symmetric spin tensor $k_{{\alpha\beta }}$ is
the derivative of a scalar field
\begin{eqnarray} \label{hsSEqS6}
k_{{\alpha\beta }}= \partial_{\alpha\beta}b(X) \; .
\end{eqnarray}
Then eqs. (\ref{hsSEqS5}) and (\ref{hsSEqS6}) reduce to the
equation (\ref{hsEqb0}) for the scalar field $b(X)$.

On the other hand, the form of the superfield (\ref{Psi=1+2}) with
$k_{{\alpha\beta }}= \partial_{\alpha\beta}b(X)$ implies that
$\Psi_{\alpha}$ is the derivative of a scalar superfield $\Phi$
obeying eqs. (\ref{hsSEq}),
\begin{eqnarray} \label{Psi=}
i\Psi_\alpha (X, \theta ) = D_\alpha \Phi (X, \theta ) \; , \qquad
D_{[\alpha } D_{\beta ] }  \Phi (X, \theta ) = 0 \; .
\end{eqnarray}

Eqs. (\ref{Psi=}) provide the general solution of eqs.
(\ref{hsSEqS1}) and (\ref{hsSEqS2}). Thus both, the scalar and
spinor superfeild representation of the system of the free higher
spin equations (\ref{hsEqb0}), (\ref{hsEqf0}) are completely
equivalent.

\def\theequation{B.\arabic{equation}}
\setcounter{equation}0
\section*{Appendix B. A generic form of the scalar superfield equation in a supergravity background}

Consider the equation
\begin{equation}\label{DDRX} {\cal D}_{[\beta}{\cal
D}_{\gamma]}\,\Phi={i\over 2}\,{\cal X}_{\beta\gamma}\, ,
\end{equation}
where ${\cal X}_{\beta\gamma}(Z)$ is an antisymmetric tensor
superfield. In Section 7 we dealt with ${\cal
X}_{\beta\gamma}=R_{\alpha\beta}\,\Phi$, and now we shall consider
the case of a generic ${\cal X}_{\beta\gamma}(Z)=-{\cal
X}_{\gamma\beta}(Z)$.

Acting on (\ref{DDRX}) with ${\cal D}_\alpha$ we arrive at a more
general form of the equation (\ref{fsg})
\begin{eqnarray}\label{fsgX}
{\cal D}_{\a[\b}{\cal D}_{\g]}\Phi = {1\over 3}R_{\b\g} \, {\cal
D}_\a\Phi - {1\over 3}R_{\a[\b} \, {\cal D}_{\g]}\Phi + {1\over
6}{\cal D}_{\alpha} {{\cal X}}_{\beta\gamma} - {1\over 6}{\cal
D}_{[\beta} {{\cal X}}_{\gamma]\alpha} \; ,
\end{eqnarray}
Then acting on (\ref{fsgX}) with ${\cal D}_\delta$ we get a
generalization of the equation (\ref{bsg})
\begin{eqnarray}\label{bsgX}
{\cal D}_{\a[\b}{\cal D}_{\g]\d}\Phi&=& {1\over 2} {\cal
D}_{\a[\b} {\cal X}_{\g]\d} - {i\over 6} {\cal D}_{\d} {\cal
D}_{\a} {\cal X}_{\b\g} + {i\over 6}  {\cal D}_{\d} {\cal D}_{[\b}
{\cal X}_{\g]\a} +
{1\over 2} R_{\a\d} {\cal X}_{\b\g} + \nonumber \\
&+&  U_{[\b\; \g]\a} \, {\cal D}_{\d} \Phi + {i\over 2} F_{\d\, \a
[\b} {\cal D}_{\g ]} \Phi + {i\over 6} {\cal D}_{\d}  R_{\b\g}\,
{\cal D}_{\a} \Phi -  {i\over 6} {\cal D}_{\d}  R_{\a[\b}\,
{\cal D}_{\g ]} \Phi
\nonumber \\
&+& i \,   {\cal D}_{\a} {\cal D}_{[\beta } \Phi\, R_{\g]\d } -
{i\over 3} R_{\b\g}\, {\cal D}_{\d}{\cal D}_{\a} \Phi + {i\over 3}
 {\cal D}_{\d}{\cal D}_{[\b} \Phi\, R_{ \b ]\a}\; .
\end{eqnarray}
The integrability condition (\ref{compare1}) of eq. (\ref{bsgX})
imposes the following restriction on the form of ${\cal
X}_{\alpha\beta}$
\begin{eqnarray}\label{selfcX}
&\hspace{-200pt}
 3{\cal D}_{(\a [\b} {\cal X}_{\g]\d)} + {\cal
D}_{\a \d} {\cal X}_{\b\g} + i {\cal D}_{(\a}{\cal D}_{ [\b} {\cal
X}_{\g]\d)} +
  4 R_{(\a [\b}
{\cal X}_{\g]\d)} \nonumber
\\
& {\hspace{70pt}}= i{\cal D}_{(\a}R_{\d)[\b}{\cal
D}_{\g]}\Phi-i{\cal D}_{[\b}R_{\g](\a}{\cal D}_{\d )}\Phi
+R_{\b\g} {\cal D}_{\a \d}\Phi + 2 R_{(\a [\b}{\cal D}_{\g]\d)}
\Phi \; .
\end{eqnarray}

One of the solutions of  (\ref{selfcX}) considered in Section 4 is
${\cal X}_{\alpha\beta}=R_{\alpha\beta}\Phi$.

As has been mentioned in Section 8, in the case of $n=2$ which
corresponds to $N=1$, $D=3$ supergravity coupled to a scalar
superfield the integrability condition (\ref{compare1}) valid for
a generic $R$ and $U$ satisfying the off--shell $SL(2)$ holonomy
constraints allows to choose ${\cal
X}_{\alpha\beta}=\epsilon_{\alpha\beta}\,f(Z)$, and in particular
${\cal X}_{\alpha\beta}=0$ or ${\cal
X}_{\alpha\beta}=m\epsilon_{\alpha\beta}$, where $m$ is a constant
of the dimension of mass.

In the case of a generic $n>2$ the scalar field equation
(\ref{DDRX}) with ${\cal X}_{\alpha\beta}=0$ is satisfied if the
right hand side of (\ref{selfcX}) vanishes. A particular solution
of this constraint is when
\begin{equation}\label{fW} \Phi=f(W)\,,
\quad  {\cal D}_{[\a}\,{\cal D}_{\b]}\Phi={\cal D}_{[\a}\,{\cal
D}_{\b]}f(W)=0\,
\end{equation}
and the superspace is superconformally flat
(\ref{DDRg})--(\ref{U=DW+WDW}) (or equivalently (\ref{e}) --
(\ref{UWf})) such that, in virtue of (\ref{fW}),
\begin{equation}\label{flatR}
R_{\a\b}=i({1\over 2}+{{f''}\over{f'}}){\cal D}_\a\,W\,{\cal
D}_\b\,W, \quad f'(W)\equiv{{df(W)}\over{dW}}\,.
\end{equation}
 In the basis of the flat covariant derivatives the scalar
superfield equation takes the form
\begin{equation}\label{flat1}
D_{[\a}D_{\b]}\,W=-(1+{{f''}\over{f'}})D_\a W\, D_\b W\,.
\end{equation}
Upon the following field redefinition  $\tilde\Phi(W)=
c\cdot\int\,e^{W } df(W)$, it reduces to the free scalar
superfield equation
\begin{equation}\label{freeg}
D_{[\a}D_{\b]}\,\tilde\Phi=0\,.
\end{equation}

{A question is whether  for $n>2$ there exist a scalar superfield
equation with ${\cal X}_{\alpha\beta}(Z)\not=R_{\alpha\beta}\Phi$
whose integrability conditions do not reduce the tensorial
superspace to the (superconformally) flat or $OSp(1|n)$
supermonifold and therefore do not trivialize it. }

\def\theequation{C.\arabic{equation}}
\setcounter{equation}0
\section{Appendix C: Peculiarities of $N=1$, $D=3$
supergravity}

Since the $N=1$, $D=3$ supergravity is a particular example of our
generic construction, before concluding the paper let us briefly
discuss this well known case from our perspective.

 In this case the holonomy group is $SL(2)\sim
Sp(2)$, and the antisymmetric tensors are proportional to
$\epsilon_{\alpha\beta}=-\epsilon_{\beta\alpha}$
$(\epsilon_{12}=1)$, for instance
$R_{\alpha\beta}(Z)=\epsilon_{\alpha\beta}R(Z)$. As one can check
(see Appendix B for details), in addition to eq. (\ref{DDR}) such
a simplification allows for other, well known, forms of the scalar
superfield equations coupled to the off--shell $N=1$, $D=3$
supergravity satisfying the constraints (\ref{Tsl1}) and
(\ref{Tsl2}), namely, the massless superfield equation
\begin{equation}\label{fsgml}
{\cal D}_{[\alpha}{\cal D}_{\beta]}\,\Phi={1\over
2}\epsilon_{\a\b}\,\epsilon^{\g\d}\,{\cal D}_\g {\cal D}_\d \Phi
=0  \quad \Rightarrow \quad {\cal D}^\alpha {\cal D}_\alpha
\Phi=0\,,
\end{equation}
and the massive superfield equation
\begin{equation}\label{fsgm}
{\cal D}_{[\alpha}{\cal D}_{\beta]}\,\Phi- {{im}\over
2}\epsilon_{\a\b}\,\Phi=\epsilon_{\a\b}({\cal D}^\g {\cal D}_\g
\Phi-im\Phi)=0 \quad \Rightarrow  \quad {\cal D}^\alpha {\cal
D}_\alpha \Phi=im\Phi\,.
\end{equation}
Moreover, in the case of $N=1$, $D=3$ superspace the non--linear
equation of the scalar superfield has the following general form
\begin{equation}\label{general}
{\cal D}^\alpha {\cal D}_\alpha \Phi=if(Z)\,\Phi\,.
\end{equation}
with $f(Z)$ being an arbitrary superfield.

Since on the mass shell $D=3$ supergravity is completely
determined by its coupling to the matter fields, we can assume
$R(Z)$ to be a function of $\Phi(Z)$, i.e. $ R=R(\Phi(Z))$. Then
(\ref{fsgml}) -- (\ref{general}) describe a non--linear
self--interaction of the scalar superfield $\Phi(Z)$.

The equations (\ref{fsgml}) -- (\ref{general}) are compatible both
with Poincare and $AdS$ $N=1$, $D=3$ supergravity. However, this
is not the case for tensorial supergravity with a generic $n$, in
which case, for example, the $Sp(n)$ holonomy required by
(\ref{fsgml}) -- (\ref{general})  reduces the tensorial
supergravity down to the supergroup manifold $OSp(1|n)$, since
${\cal D}C_{\alpha\beta}=0$.

Let us note that in the case of $N=1$, $D=3$ supergravity  the
equations (\ref{DDRW})--(\ref{U=DW+WDW}), (\ref{dynamic}) and
(\ref{dynamico}) (with a generic function of $W$ (or $\Phi$) on
the right hand side), or equivalently eqs. (\ref{general}), are
Lagrangian in the sense that they can be derived from the $N=1$,
$D=3$ supergravity action \cite{D=3} coupled to a scalar field
\begin{equation}\label{D=3}
S={1\over 2}\,\int \,d^3x d^2\theta ~{\rm sdet}\,
E^{~A}_B\left[R+\epsilon^{\a\b}{\cal D}_\a \Phi {\cal D}_\b
\Phi+\varsigma\cdot {\cal X}(\Phi)\right]\,.
\end{equation}

In the generic case of $n>2$ it is still an open problem to figure
out whether the equations (\ref{DDRW})--(\ref{U=DW+WDW}),
(\ref{dynamic}) and (\ref{dynamico}), as well as (\ref{hsSEq}) and
(\ref{DD=AdS}) can be obtained by the variation of a corresponding
action.

\end{document}